\documentclass[]{aa}  

\usepackage{graphicx}
\usepackage{subfig}
\usepackage[varg]{txfonts}
\usepackage{morefloats} 
\usepackage[utf8]{inputenc}
\usepackage{ textcomp }
\usepackage{ gensymb }  
\usepackage{float}
\usepackage{multirow}
\usepackage{blindtext}
\usepackage{enumitem}
\usepackage{scrextend}
\usepackage{natbib,twoopt}
\usepackage{url}
\usepackage[breaklinks=true]{hyperref} 
\usepackage{color}
\usepackage{soul}

\def\halpha{H$\mathrm{\alpha}$}
\def\kms{\hbox{km$\;$s$^{-1}$}}
            
\begin{document}

   \title{Multi-scale observations of thermal non-equilibrium cycles in coronal loops}

   \author{C. Froment\inst{1,2}\thanks{Present address: LPC2E, CNRS and University of Orl\'eans, 3A avenue de la Recherche Scientifique, Orl\'eans, France}%[0000-0001-5315-2890]
          \and
          P. Antolin\inst{3}%[0000-0003-1529-4681]
          \and
          V. M. J. Henriques\inst{1,2}%[0000-0002-4024-7732]
          \and
          P. Kohutova\inst{1,2}%[0000-0001-9213-7773]
          \and
          L. H. M. Rouppe van der Voort\inst{1,2} %[0000-0003-2088-028X]
          }

   \institute{Rosseland Centre for Solar Physics, University of Oslo, P.O. Box 1029 Blindern, NO-0315, Oslo, Norway \\ \email{clara.froment@cnrs-orleans.fr}
   		  \and
	Institute of Theoretical Astrophysics, University of Oslo, P.O. Box 1029 Blindern, NO-0315, Oslo, Norway
    	  \and
    Department of Mathematics, Physics and Electrical Engineering, Northumbria University, Newcastle upon Tyne, NE1 8ST, UK
             }
\date{Received 17 September 2019; Accepted 18 November 2019}

\abstract
 {Thermal non-equilibrium (TNE) is a phenomenon that can occur in solar coronal loops when the heating is quasi-constant and highly-stratified. Under such heating conditions, coronal loops undergo cycles of evaporation and condensation. The recent observations of ubiquitous long-period intensity pulsations in coronal loops and their relationship with coronal rain have demonstrated that understanding the characteristics of TNE cycles is an essential step in constraining the circulation of mass and energy in the corona.}
{We report unique observations with the Solar Dynamics Observatory (SDO) and the Swedish 1-m Solar Telescope (SST) that link the captured thermal properties across the extreme spatiotemporal scales covered by TNE processes.}
 {Within the same coronal loop bundle, we captured 6 hr period coronal intensity pulsations in SDO/AIA and coronal rain observed off-limb in the chromospheric \halpha\ and \ion{Ca}{II}~K spectral lines with SST/CRISP and SST/CHROMIS. We combined a multi-thermal analysis of the cycles with AIA and an extensive spectral characterisation of the rain clumps with the SST.} {We find clear evidence of evaporation-condensation cycles in the corona which are linked with periodic coronal rain showers. The high-resolution spectroscopic instruments at the SST reveal the fine-structured rain strands and allow us to probe the cooling phase of one of the cycles down to chromospheric temperatures.}{These observations reinforce the link between long-period intensity pulsations and coronal rain. They also demonstrate the capability of TNE to shape the dynamics of active regions on the large scales as well as on the smallest scales currently resolvable.}


   \maketitle
%
%________________________________________________________________

\section{Context}\label{sec:intro}

Thermal non-equilibrium (TNE) can explain several facets of the dynamic solar corona \citep{antiochos_model_1991, Muller_2003AA...411..605M}, yet recent works \citep[e.g.][]{Antolin_2015ApJ...806...81A, froment_evidence_2015, auchere_coronal_2018} suggest that TNE could be far more prevalent than previously thought.
This phenomenon can occur in coronal loops when they are heated predominantly at their footpoints, quasi-steadily (i.e. high frequency of repetition of the heating events compared to the radiative cooling time). Counter-intuitively, under these very stable heating conditions, the loops are not able to maintain a thermal equilibrium. On the contrary, their behaviour alternates between heating and cooling phases in which  a loop bundle is first filled by evaporative-chromospheric plasma and then partially drained when condensations form and are evacuated through the footpoints. If the heating conditions remain stable or quasi-stable, it results in evaporation and condensation cycles \citep{kuin_thermal_1982},  whose period depends on the length and geometry of the structures and the heating characteristics \citep{froment_occurrence_2018}.
The formation of the condensations in the corona is initiated by an excess of density. The radiative losses, increasing with the density and the decrease of the temperature at coronal conditions, outweigh the energy input \citep[see e.g.][for more details on the TNE mechanism]{klimchuk_role_2019}, triggering a thermal runaway (or catastrophic cooling) once the temperature decreases below 1~MK. As such, the characteristics of the TNE cycles reflect the details of the spatio-temporal distribution of the heating \citep[e.g.][]{johnston_effects_2019}.

This thermal runaway has all the characteristics of a thermal instability \citep{Parker_1953ApJ...117..431P, Field_1965ApJ...142..531F}. However, in TNE conditions there is by definition no equilibrium. Semantically, thus, we cannot formally talk of a thermal instability. On the other hand, thermal instabilities can play a role in loops marginally in equilibrium. Furthermore, considering that the scales of TNE and condensations formation are fundamentally different, we could probably talk of a local thermal instability to describe the runway cooling that leads to the formation of condensations during TNE cycles. Further details on this matter can be found in Klimchuk (under revision) and  \citet{antolin_thermal_2019}.

The role of TNE in explaining the widespread existence of cool material at coronal heights has been largely explored. It is the cases for prominences \citep[e.g.][]{antiochos_model_1991, Antiochos_2000ApJ...536..494A, karpen_origin_2006, xia_formation_2011, Xia_2016ApJ...823...22X}, when the condensations have accumulated or coronal rain \citep{schrijver_catastrophic_2001,Muller_2003AA...411..605M,DeGroof_2004AA...415.1141D,Muller_2004AA...424..289M, antolin_coronal_2010, fang_multidimensional_2013, Moschou_2015AdSpR..56.2738M} when the condensations are evacuated. Whatever its global structural form, the physical process responsible for the presence of these cool ($10^4$ to $10^5$~K) and dense ($10^{10}$ to $10^{12} \; \mathrm{cm}^{-3}$) condensations in the corona, is believed to be similar \citep{Antiochos_1999ApJ...512..985A}.

Coronal rain was first reported by \citet{Kawaguchi_1970PASJ...22..405K} and \citet{Leroy_1972SoPh...25..413L}. 
High-resolution and space-based observations later revealed that coronal rain is in the form of blob-like structures with widths of a few hundreds of km, observed in chromospheric and transition-region lines \citep{Antolin_2015ApJ...806...81A}. These blobs appear spatially organised in clumps with velocities of up to 100-150~\kms\ and temporally organised in different episodes called \lq showers\rq\,in \citet{Antolin_Rouppe_2012ApJ...745..152A}. Easily visible within coronal loops extending off-limb, coronal rain seems to be ubiquitous in active regions.

More recently, the coronal signatures of TNE were detected in the form of long-period EUV pulsations (2-16 hours) that appeared to be very common in the solar corona and in particular in coronal loops \citep{auchere_long-period_2014, 2016PhDT.......115F}. 
Detailed analyses with the Atmospheric Imaging Assembly \citep[AIA,][]{boerner_initial_2012, lemen_atmospheric_2012} on board the Solar Dynamics Observatory \citep[SDO,][]{pesnell_solar_2012} in comparison with 1D hydrodynamic simulations, showed that the characteristics of these pulsating loops match the one predicted by the TNE model \citep{froment_evidence_2015, froment_long-period_2017}.
Moreover, the characteristics of their signal (shape of the Fourier spectra) support the TNE explanation \citep{auchere_thermal_2016}.

The analysis presented in \citet{auchere_coronal_2018} eventually showed that these EUV pulsations are undoubtedly the coronal  manifestation of the TNE cycles while coronal rain (its transition-region and chromospheric alter ego) appear periodically in the 304~\AA~channel of SDO/AIA during its cooling phases. This study also pointed out the widespread occurrence of coronal rain within the active regions observed, whether periodic or not, implying that quasi-constant footpoint heating is playing a role on a large scale. To the best of our knowledge, if the periodicity of the evaporation-condensation cycles can only be explained by TNE, an absence of periodicity does not necessarily imply a different physical explanation  than quasi-constant footpoint heating. First of all, a lack of periodicity can be explained by heating and geometry conditions varying significantly over what would be a period \citep{antolin_coronal_2010}. The varying heating conditions can even break the generation of new cycles if they drift away from TNE-prone conditions. Additionally, the intrinsic nature of the detection technique used, and line-of-sight (LOS) effects, are likely narrowing down the number of detections of periodic coronal rain and periodic EUV intensity pulsations.

It is now clear that TNE is playing a role at a large scale in active regions. However, proper statistics of the fraction of the coronal volume undergoing TNE (regardless the type of structure involved or the temperature and the density reached) are still lacking. Its quantitative role in the mass cycle and heating of the solar atmosphere has yet to be determined. Recent works have shown that apart from stratified heating in the atmosphere leading to TNE and, thus, evaporation-condensation cycles, other ways may exist to reach this rain-triggering thermal instability: magnetic reconnection between open-structure and low-lying closed loops \citep{Liu_2016SPD....47.0402L, li_coronal_2018}, interchange reconnection in null-point topologies \citep{mason_observations_2019}, and reconnection-induced localised heating \citep{kohutova_formation_2019}. 
Due to their recent discovery, the details of the mechanisms that take place in such configurations are currently unclear. 
Furthermore, coronal rain also seems to be always present in flaring loops \citep[e.g.][]{Scullion_2014ApJ...797...36S, Scullion_2016ApJ...833..184S}. However, it is unclear how the heating from the electron beams acts to produce coronal rain \citep{reep_electron_2018}.
If all these types of coronal rain events need to be understood to draw a comprehensive picture of the role of thermal instability and thermal runaway in the solar atmosphere, it is clear that their transient nature\footnote{Except for the coronal rain in null-point typologies that are also long-lived structures.} make them very different from the so-called quiescent coronal rain, which is recurrent in active regions. 

The lack of thermal equilibrium consequent to the concentration of the heating in the lower portions of the coronal loops, was predicted early by \citet{Serio_1981ApJ...243..288S} along with the periodic evolution of temperature and density of the plasma resulting from such heating configurations \citep{kuin_thermal_1982}. 
In recent years, a significant modelling effort has been deployed in order to understand the role of TNE in shaping the global dynamics of active regions \citep[e.g.][]{mok_calculating_2005, lionello_thermal_2013, mikic_importance_2013,Fang_2013ApJ...771L..29F,Fang_2015ApJ...807..142F,Moschou_2015AdSpR..56.2738M, mok_three-dimensional_2016,Xia_2017AA...603A..42X}, as well as prominences \citep[e.g.][]{xia_formation_2011,Luna_etal_2012ApJ...746...30L,xia_simulating_2014, Xia_2016ApJ...823...22X}.

One central challenge is the determination of observables that could help discriminate between quasi-constant footpoint heating and non-localised heating \citep{klimchuk_can_2010,lionello_thermal_2013,winebarger_verification_2014,lionello_can_2016,Winebarger_2016ApJ...831..172W,Winebarger_2018ApJ...865..111W}. In this context, it is crucial to look for the various TNE configurations predicted by the models. In coronal loops whose global behaviour is ruled by TNE, the observables appear at many different scales, namely, temporal: cycles that can go on for days vs the rain evolving over a few minutes; spatial: coherent loop evolution over hundreds of Mm vs rain widths on the order of hundreds of km; and thermodynamics: from coronal to chromospheric plasma characteristics. Observational analyses covering this wide range of different spatiotemporal scales ought to be carried out but have been missing up to now.

In this paper, we present a detailed analysis of the thermal evolution of the external loops of the NOAA AR 12674 active region which seem dominated by TNE. The northern footpoint of these loops was observed off-limb on August 29, 2017 with the Swedish 1-m Solar Telescope \citep[SST,][]{scharmer_1-meter_2003}. Using a combination of these high-resolution ground-based observations and data from SDO/AIA, we were able to follow both the coronal response of TNE cycles in the form of long-period EUV pulsations and the fine coronal rain structures that appeared in chromospheric lines during the cooling phase of one of the cycles.  After a presentation of the AIA and SST datasets, the periodic intensity pulsations and coronal rain showers observed in Sect.~\ref{sec:data}, we derive the physical characteristics of this event. The first part is a study of the coronal cooling cycles, via a Differential Emission Measure (DEM) and a time-lag analysis, presented in Sect.~\ref{sec:cycles_analysis}. The second part, that is, the analysis of the fine rain structures and their physical properties, is detailed in  Sect.~\ref{sec:sst_analysis}. From this work we are thus able to extract the key observational properties of TNE cycles with complete condensations. This is important for future comparisons with numerical simulations and, therefore, constraining the solar atmospheric heating models.

\section{Long-period intensity pulsations and coronal rain with SDO and SST observations}\label{sec:data}

\subsection{AIA observations} 

\subsubsection{Data sample}\label{sec:aia_data}

Using a modified version of the detection algorithm presented in \citet{auchere_long-period_2014}, we followed a region of interest (ROI) located at the East limb from August 28, 2017 00:00 UT to August 30, 2017 00:00 UT. As already mentioned in \citet{auchere_coronal_2018}, the main difference with regards to the on-disk version of this algorithm is that we had to turn-off the mapping into heliographic coordinates because of the distortions of off-limb structures caused by this coordinate transformation. For the present paper and in order to facilitate the combined analysis with the SST data, we chose to keep the SDO images in Cartesian coordinates. 
Mapping into polar coordinates was not needed since the studied active region is much smaller than the one showing trans-equatorial loops in \citet{auchere_coronal_2018}.

The AIA observational sequence was limited to a three-day interval in order to minimise the height variations of the structures off-limb due to the rotation of the Sun. Even if we did not observe the loops of interest being at the same spatial position during the entire sequence, we chose not the reduce the length of observations in order to keep a reasonable frequency resolution for the Fourier analysis (i.e. $3.84 \; \mathrm{\micro Hz}$ per frequency bin).

\begin{figure}
    \centering
	\includegraphics{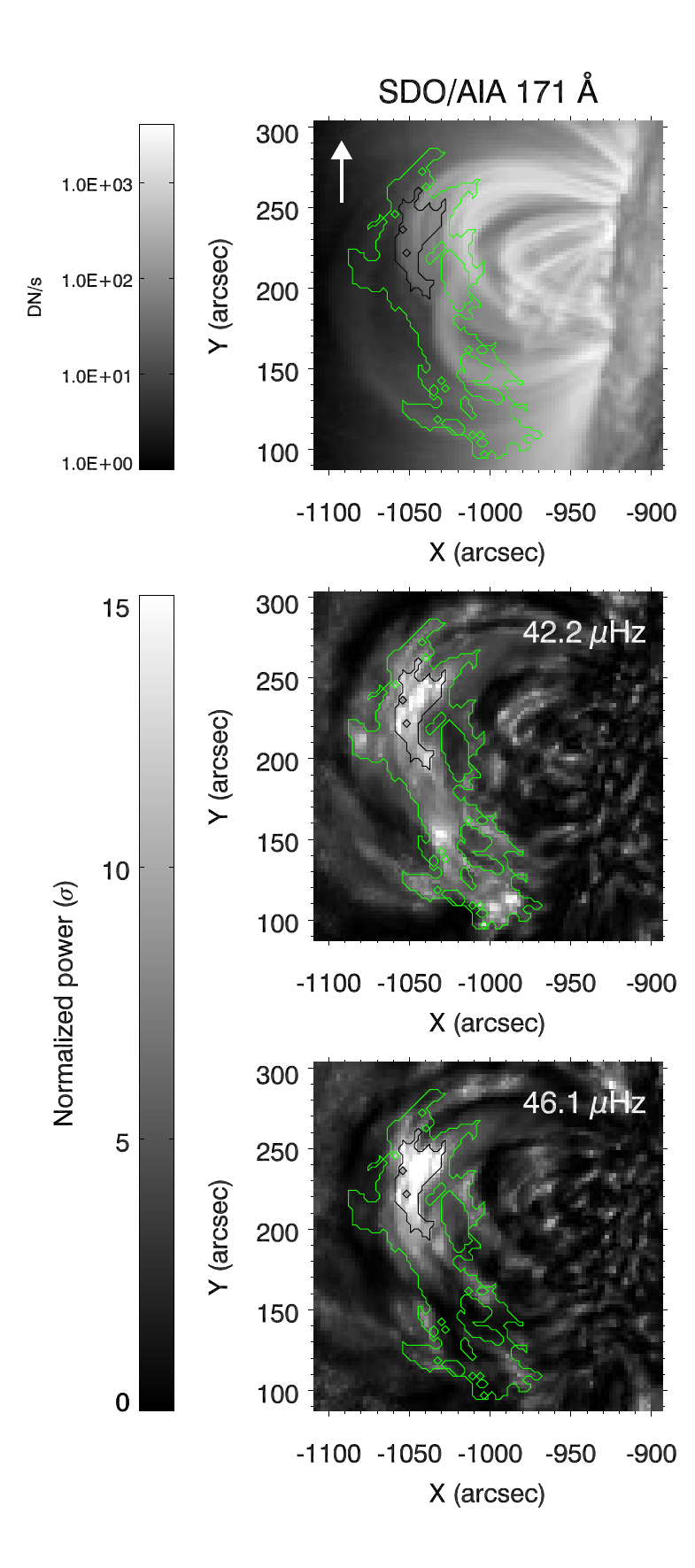}
	\caption{Top: 171~\AA~image at August 29, 2017 08:49:45 of the SDO/AIA ROI just before the beginning of the SST dataset. The green and black contour represent, respectively, the regions detected with normalised power above $5\sigma$ (extended contour) and $10\sigma$. The white arrow indicates the north direction.
	Middle: normalised power map at 42.2~$\mu\mathrm{Hz}$ (i.e. 6.6~hours) for the 171 time series.
	Bottom: normalised power map at 46.1~$\mu\mathrm{Hz}$ (i.e. 6.0~hours) for the 171 time series. The scale is saturated to $15\sigma$ for both frequency bands. 
	}
	\label{fig:img_aia_power_maps}
\end{figure}

The ROI is presented in the top most panel of Fig.~\ref{fig:img_aia_power_maps} in the 171~\AA~passband of AIA on August 29, 2017 at 08:49:45~UT, that is, just before the beginning of the SST observations.
The ROI is $216^{\prime\prime} \times 216^{\prime\prime}$ and centred at (X,Y) = ($-1001.3^{\prime\prime}$, $195.3^{\prime\prime}$).
The AIA data are read and calibrated to level 1.5 using the routine \texttt{aia\_prep} from the Interactive Data Language SolarSoftware library. As in \citet{froment_evidence_2015} we used a cadence of 13 minutes and normalise the intensities to the exposure times.
We also corrected for some jitter seen at several times during the sequence by fitting the limb and determining the new disk-centre position. We note that the sequence has some data gaps, in particular between approximately 6:30 and 7:30 UT on August 29, due to the eclipse season.

For the detection of the periodic signal, the AIA images are binned $4\times4$ in order to increase  the signal-to-noise ratio.
However, for the rest of the analysis presented in this paper (from Sect.~\ref{sec:cycles_analysis}), we used the full resolution provided by AIA. Moreover, we used the 12 second cadence for co-alignment with SST observations and 1 minute cadence for the DEM and time-lag analysis.
As in \citet{froment_evidence_2015}, we used the six coronal channels of AIA: 94~\AA, 131~\AA, 171~\AA, 193~\AA, 211~\AA, 335~\AA, to which we add the 304~\AA~channel that has both a coronal component around 1.8~MK and a transition region component around 0.08~MK (peaks in the temperature response of the passband).

\begin{figure}
	\centering
	\includegraphics[width=\hsize]{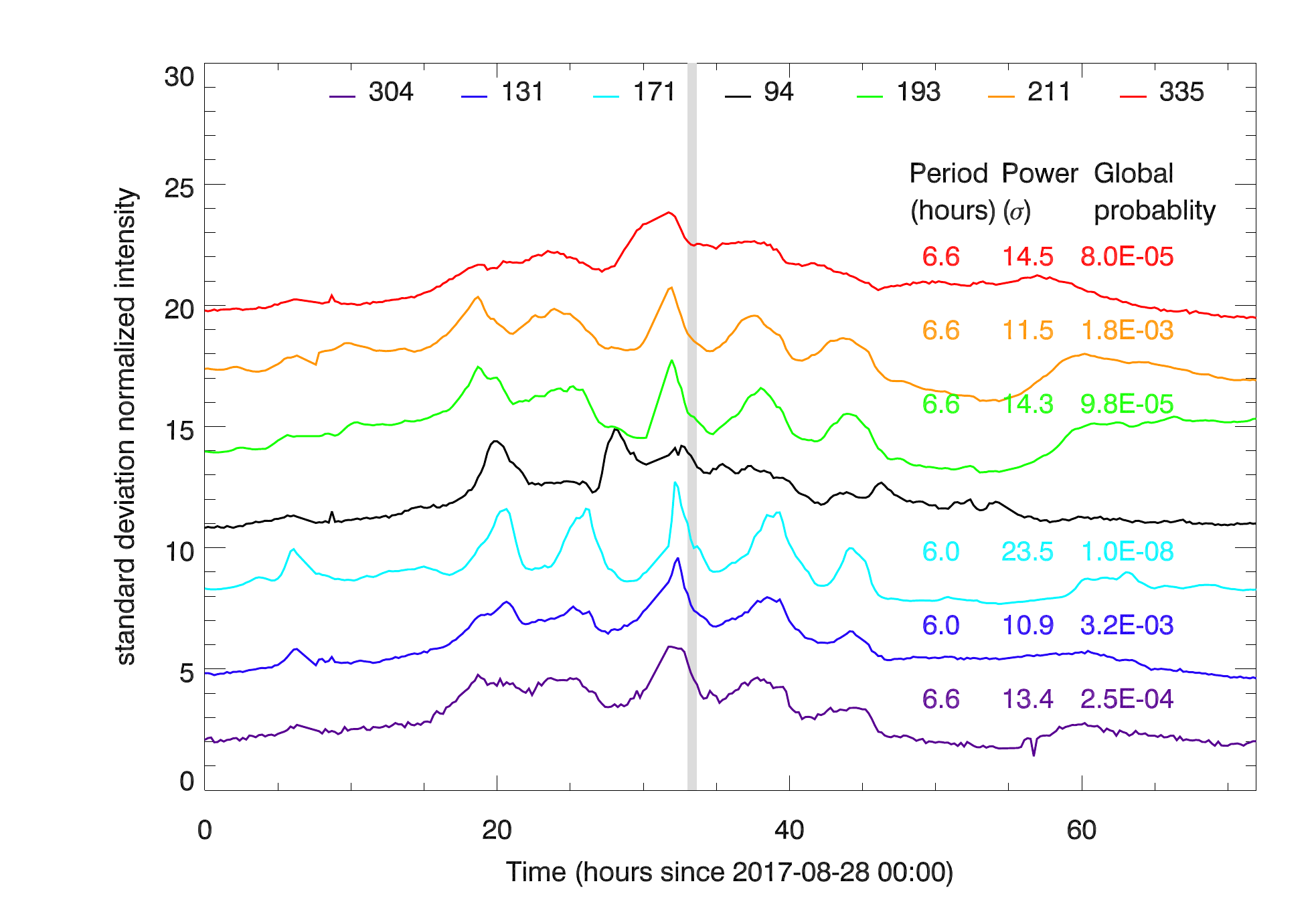}
	\caption{Evolution of the intensity for seven AIA channels for three days, starting at 2017-08-28 00:00 UT. These time series are averaged over the black contour displayed in Fig.~\ref{fig:img_aia_power_maps}. All the light curves are normalised to standard deviation and offset by arbitrary numbers on the y-axis. For each curve, we display the corresponding period, Fourier power and the associated random occurrence probability. The grey bar indicates the duration of the SST observations.
	}
	\label{fig:light_curves_AIA}
\end{figure}

\subsubsection{Periodic pulsations analysis}\label{sec:period_analysis}

The power spectral density (PSD) is computed for each time series corresponding to each pixel present in the ROI. This means that we did not try to look for pulsations in a particular region of the ROI, nor try to follow a particular structure. Only the final area of excess of power that we analyse for each frequency bin between $18.5~\mu\mathrm{Hz}$ and $110~\mu\mathrm{Hz}$ was used to identify the underlying structures in the AIA images.
We used global confidence levels \citep{auchere_fourier_2016} to determine the significance of the power peaks against a local estimate of the noise $\sigma(\nu)$.
The global probability that at least one frequency bin among $N/2$ shows an excess of power $m$ times greater than $\sigma(\nu)$ is expressed as follows: $P_g(m)=1- (1 - \mathrm{e}^{-m})^{N/2}$, with $N=332$ the number of time steps for each time series.
This procedure is introduced in \citet{auchere_long-period_2014} and used in \citet{froment_evidence_2015, auchere_coronal_2018}, for example. For more details about the detection algorithm, we refer the readers to these papers.

In the middle and bottom panels of Fig.~\ref{fig:img_aia_power_maps}, we show two normalised power maps for the field of view, computed for the 171~\AA~channel of AIA (i.e. the passband for which we found the strongest signal). These two maps are, respectively, for 42.2~$\mu\mathrm{Hz}$ (i.e. 6.6~hours) and 46.1~$\mu\mathrm{Hz}$ (i.e. 6.0~hours), which are the only two frequency bands where we detect a strong signal,  above a 99~\% confidence level threshold (corresponding to 10~$\sigma$), with sufficient spatial coherence\footnote{The size of the final contour has to match criteria detailed in \citet{auchere_long-period_2014}. One pixel among all the pixels of the FOV is likely to show a significant periodic signal by chance. However, this probability decreases with an increasing number of adjacent pixels above the threshold, especially when forming spatially meaningful structures.}. 

While the black contour in all the panels of Fig.~\ref{fig:img_aia_power_maps} represents the region where the power is above the  $10\sigma$ threshold, the green contour corresponds to the extended region where the power is greater than $5\sigma$. This green contour has an area of $(923^{\prime\prime})^2$. It encompasses the apex area of the external loops of the active region, thus covering different loop bundles. The maximum power in this contour for the two frequency bands previously listed is, respectively, $19.9\sigma$ and $26.1\sigma$. The southern loop bundle, that is the extension of the contour around ($-1020^{\prime\prime}$,$160^{\prime\prime}$), seems to show a periodicity only in the 46.1~$\mu\mathrm{Hz}$ frequency band.
Our rain observations with the SST are located at the footpoints, around ($-920^{\prime\prime}$, $250^{\prime\prime}$), of the northern loop bundle, and since the signal is stronger in this  bundle (where the $10\sigma$ contour is located), we chose not to analyse in detail the southern bundle. Moreover, the power in this area is affected by the main loop bundle since it rotates into the same coordinates (see Movie~1). For the analysis presented in this paper, we focus, thus, on the smaller contour (black contour) restricted to the $10\sigma$ threshold, located at the apex of the northern bundle. The average power in the contour is, respectively, $11.3\sigma$ and $13.2\sigma$ for 42.2~$\mu\mathrm{Hz}$ and 46.1~$\mu\mathrm{Hz}$, which is, respectively, $2 \times 10^{-4}$  and $3 \times 10^{-5}$ of random occurrence probability $P_g(m)$ for the 171~\AA~passband (middle and bottom panel of Fig.~\ref{fig:img_aia_power_maps}). 

In Fig.~\ref{fig:light_curves_AIA}, we show the normalised intensity time-series averaged over this small contour for the 335, 211, 193, 94, 171, 131 and 304~\AA~channels of AIA.
We are able to detect clear pulsations for 6 out of these 7 channels.
The only passband where the normalised power does not exceed the $10\sigma$ threshold is 94. For the rest of the channels, the global probability $P_g(m)$ that these pulsations are due to noise is extremely low: between $3.2 \times 10^{-3}$ and $1.0 \times 10^{-8}$ depending on the passbands.  
The lack of detection in the 94 channel could be explained by the low signal-to-noise ratio of this passband. 
However, some pulses are visible (at least one 20 hours after the beginning of the sequence) in the time series,  albeit the Fourier analysis is not showing a periodic signal. In all the other channels, five intensity pulses can be clearly identified between 15 and 45~hours after the beginning of the sequence. We will later refer to this time period by \lq the five central pulses\rq.

\begin{figure}
	\resizebox{\hsize}{!}
	{\begin{tabular}{c} 
		\includegraphics[trim={0cm 0cm 1.3cm 1cm},clip]{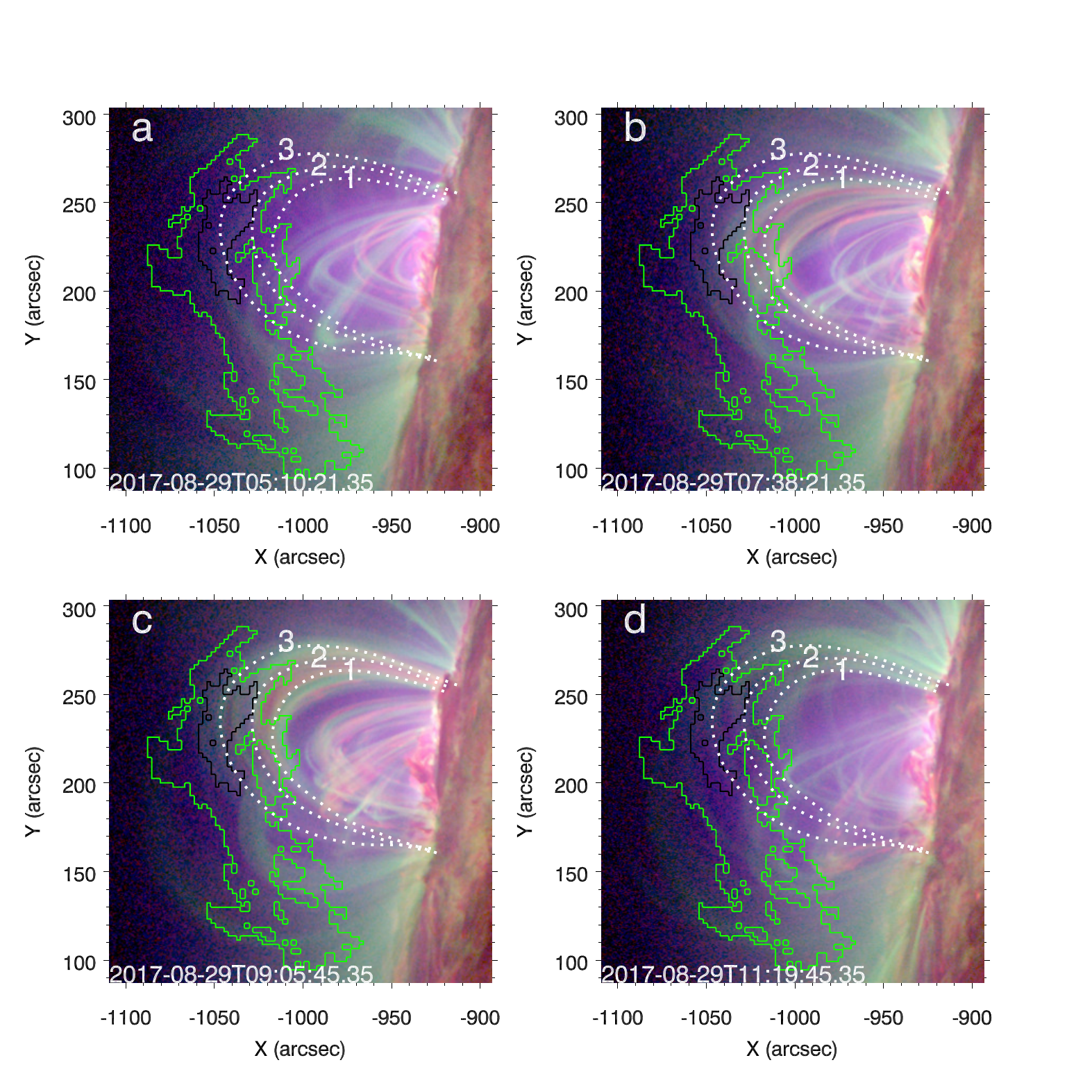} \\
		\includegraphics{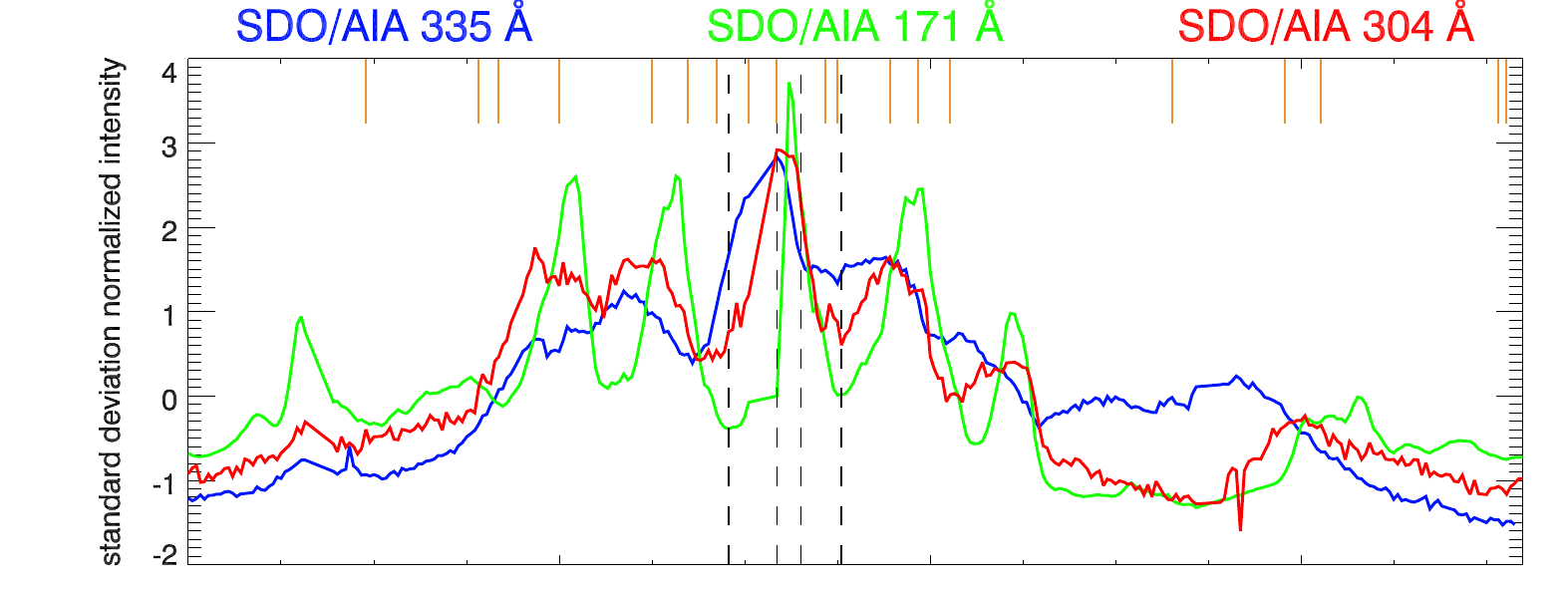} \\
        \includegraphics{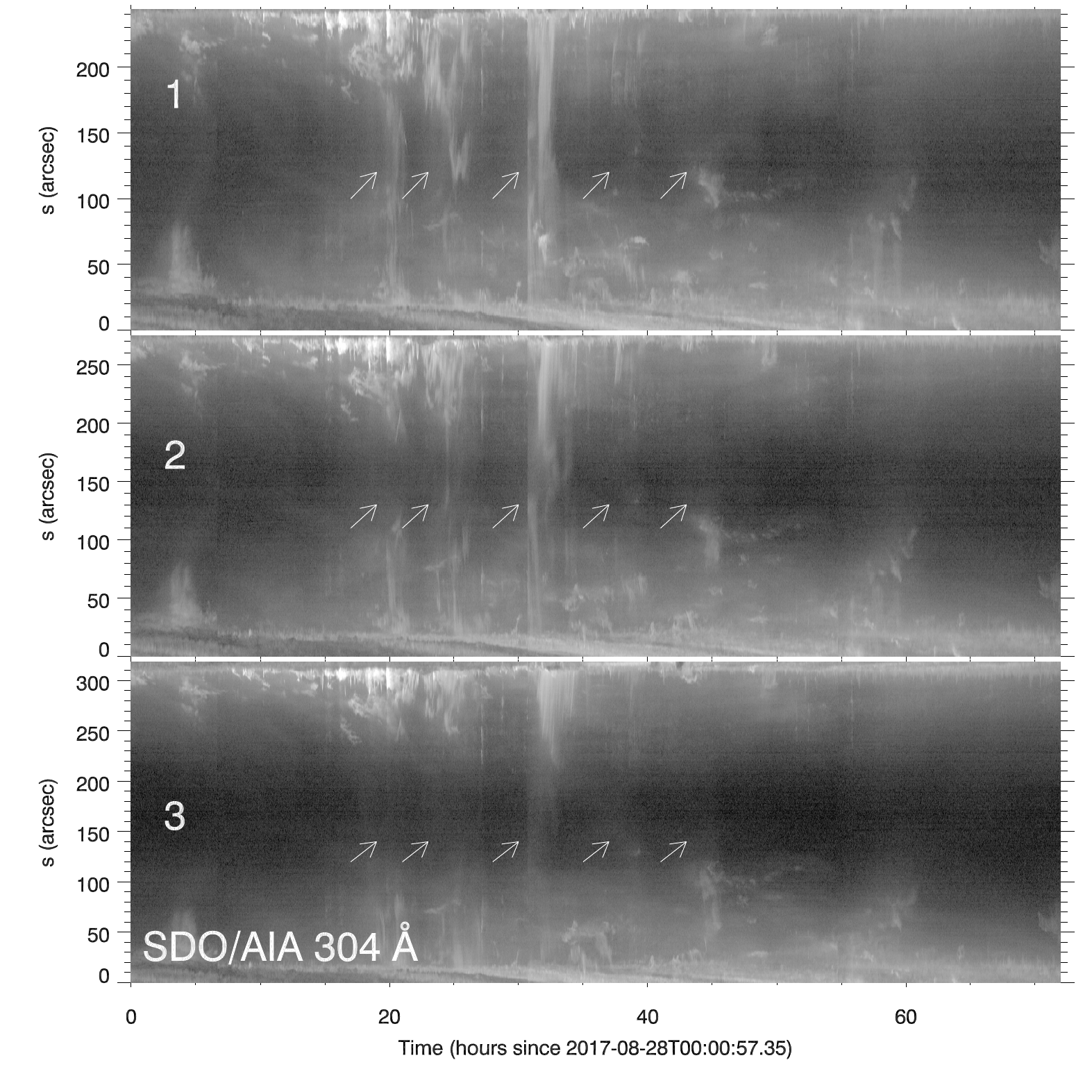}
	\end{tabular}
	}
	\caption{Periodic coronal rain and coronal emission in the 304 channel. Top: Four stages observed during the third rain cycle (details in Sect.~\ref{sec:aia_304_rain}). RGB images combining the 335~\AA~AIA channel in blue, 171~\AA~in green and 304~\AA~in red. The green and black contours are, respectively, the $5\sigma$ and $10\sigma$ contours of detection. Middle: Evolution of the intensity for the same three channels (same light curves as in Fig.~\ref{fig:light_curves_AIA}). The four vertical dotted black lines mark the time of the four images of the figure. The small orange vertical solid lines mark the approximate time for some of the flares occurring during the sequence. Bottom: Time-distance plots taken along the three dotted lines shown in the top panel. The origin of the y-axes is the northern footpoint of the lines. The white arrows points out the five cycles. (An animation of this figure is available.)}
	\label{fig:rain_cycle}
\end{figure}

\subsubsection{Recurring coronal rain showers}\label{sec:aia_304_rain}

In the 304~\AA~channel of AIA, coronal rain is seen falling regularly toward both legs of the loop bundle showing EUV pulsations, mostly in phase with the decreasing phase of the intensity cycles (Movie~1  and middle and bottom panels of Fig.~\ref{fig:rain_cycle}). 
In Fig.~\ref{fig:rain_cycle}, we show different stages observed during the third intensity cycle, during which we observed with the SST. The top panel shows the 335, 171 and 304~channels evolution with RGB images taken at four different stages of this cycle. The time of each stage is indicated on the time series in the middle panel of the figure so the rain formation in space and time can be followed. In panel~a, the pulsating loop bundle is barely visible, it is the minimum in intensity preceding the start of the cycle. In panel~b, the rain event starts in the inner edge of the bundle, where the smaller loops of the large pulsating bundle can be found. The rain appears to form mostly at their apex. In panel~c (at the start the SST observations), the rain starts forming also in the outer edge of the bundle, also at the loops apex. The rain episode goes on during the decreasing phase of the intensities and then stops. Panel~d corresponds to the post-evacuation phase, the intensity minimum is reached again.

Looking at the entire sequence, in particular during the five central pulses, it actually seems that the rain is forming mostly in the inner edge of the pulsating bundle apex rather than in its outer edge (i.e. where the 10~$\sigma$ detection contour is located). This is well illustrated by the time-distance plots made in the 304~channel and shown in the bottom panel of the same Figure. These plots were constructed taking three different loop paths within the pulsating loop bundle, covering the inner and outer edge the apex region. These loop paths were constructed to match the loops during the cycle presented in the top panel images. Because we did not compensate for the solar rotation, the match is, therefore, not perfect for the other cycles. Thus, such plots cannot be used to determine the rain trajectories for example but give a good indication of when and where the rain forms. For the inner loop (loop path~1), five rain events (pointed out by the white arrows) are simultaneous to the intensity pulses and are showing a characteristic clumpy emission starting at the loop apex. As previously stated, the loop paths do not match well with the loop bundle anymore for the two last events. From these plots only it is thus unclear where the rain starts to form.
The loop path~2 shows a similar situation as for number~1. However, for loop path~3, that is the one in the outer edge of the bundle, the 304 emission near the loop apex is rather hazy even if the clumpy rain emission can still be found in the legs. Since this area coincides with the 10~$\sigma$ detection contour, this suggests that the periodic signal detected in the 304~\AA~channel (see middle panel) may not be due to the coronal rain occurrence itself. As it is developed later in Sect.~\ref{sec:time_lag}, the $10\sigma$ signal seem to be dominated by the 1.8~MK component of 304. The pulsations detected in this channel thus seem to be mostly due to the cooling through the coronal temperatures (like for the other coronal channels) rather than the rain itself. However, the time-distance plots give a clear evidence that the rain events in this bundle are periodic.

As a final note, coronal rain still clearly forms in the same bundle where we detect long-period intensity pulsations,  well after any pulses were detectable in the light curves and until the very end of the sequence.  This is due to the fact that we did not compensate for solar rotation.  Toward the end of the sequence, the loops of interest do not coincide anymore with the black contour but it seems that they are still contained within the green contour (see Movie~1).

Another interesting characteristic for this active region is the  \lq
intensity wavefront\rq\,observed across several loop bundles, including those for which the pulsations are found (see Movie 1). This behaviour is characteritic of TNE, as revealed by global 3D MHD simulations \citep{mok_three-dimensional_2016}. Indeed, due to the continuous increase in length for loops away from the AR core, under similar heating conditions the TNE cycle period is proportionately increased. The time of peak intensity for each channel is therefore delayed for longer loops, thereby giving the apparent expansion effect.

We carefully checked that during this three-day sequence there is no flaring activity linked to the rain production in the selected loop bundle. There are several small-class flares that are observed during the three-day dataset but they always happen fairly low in the atmosphere, in a prominence-like structure that does not seem to be connected to the pulsating coronal loop bundle\footnote{However, it is possible that the flaring activity is linked to the production of some of the coronal rain in smaller loops that are located in the core of the active region.} (see Movie~1). Moreover, there are no flares above C-class \citep[i.e. as for the active regions studied in][]{froment_evidence_2015} and no apparent correlation between the rain occurrence with the flares (see times displayed with the orange dotted lines in Fig.~\ref{fig:rain_cycle} and Movie~1) neither in time nor in location. We can thus safely conclude that the coronal rain observed is not produced in a post-flare configuration.

In this active region, coronal rain is not only present in the pulsating loop bundle, but in multiple locations, from the core to the external loops.
However, we do detect long-period intensity pulsations in only one particular bundle. Since the cycle periods typically scale with the loops length, we also investigated shorter periods than 6~hours in shorter loops, such as the ones of the core of the active region. We found no significant signal in any frequency range explored.
This non-detection of periodic signal does not necessarily imply that there are no TNE cycles in these other raining loops. The core of this active region is rather complicated, many different loops with different geometries entangle along the LOS, making any periodic signal likely difficult to detect. It can also mean that, for these loops, there is no intrinsic periodicity, implying that the heating conditions are changing too much to have a regular repetition of cycles.

\subsection{SST observations}\label{sec:sst_data}

\subsubsection{Data sample}

On August 29, 2017 we followed the same active region throughout the day at the SST with the CRisp Imaging SpectroPolarimeter \citep[CRISP,][]{scharmer_crisp_2008} and the CHROMospheric Imaging Spectrometer (CHROMIS, Scharmer, in preparation) instruments. 
The observations were aided by the SST adaptive optics (AO) system, which consists of a tip-tilt mirror and an 85-electrode deformable mirror controlled by a Shack-Hartmann wavefront sensor \citep[this is an upgrade of the AO system described in][]{2003SPIE.4853..370S}. 
While good AO performance is generally difficult for off-limb observations, a near-limb high-contrast facular point served as a suitable reference AO locking point and the system successfully operated in closed-loop throughout the analysed time range. 
The dataset that we analyse in the present paper spans from 09:05:59 to 09:32:35 with CRISP, and 09:07:10 to 09:24:27 with CHROMIS.  
The vertical grey bar in Fig.~\ref{fig:light_curves_AIA} indicates the duration of the CRISP dataset.
The SST observations were taking place during the decreasing phase of one of the pulses shown in the AIA intensities.

The CRISP field-of-view (FOV, 58\arcsec $\times$ 57\arcsec) was centred at $(X,Y) = (-919\arcsec, 219\arcsec)$ 
at the beginning of the observations.
The CHROMIS FOV was more rectangular in size (65\arcsec $\times$ 38\arcsec). 
With the CRISP instrument we scanned the \halpha\ line in the spectral window $6563\pm1.6$~\AA~(equivalent to $\pm73$~\kms).
We sampled 32 line positions with mostly 0.1~\AA~steps (4.57~\kms).
The cadence, that is the time between two scans, was 7.29 seconds and the pixel size 0\farcs058.

With the CHROMIS instrument, we sampled the \ion{Ca}{II}~K line ($3 933.664~\AA$) with 41 positions in the $3933.69\pm1.3$~\AA~spectral window ($\pm99$~\kms), using 0.065~\AA~steps ($\pm5$~\kms).
One continuum point at 400~nm was added. The cadence was about 14 seconds and the pixel size 0\farcs038. 

\subsubsection{Data reduction}

The \halpha\ observations were processed following the CRISPRED reduction pipeline \citep{de_la_cruz_rodriguez_crispred:_2015}.
The CHROMIS data reduction was performed with a protoype version of the CHROMISRED pipepline \citep{lofdahl_data-processing_2018}.
The image reconstruction was performed using Multi-Object Multi-Frame Blind Deconvolution \citep[MOMFBD,][]{lofdahl_momfbd_2002SPIE.4792..146L,van_noort_solar_2005} with 80 Karhunen-Lo\'{e}ve modes sorted in order of atmospheric significance. 

The co-alignment of the CRISP and CHROMIS FOVs was performed as a single shift per scan computation using cross-correlation of an on-disk sub-field common to the wideband channels of both instruments. Some misalignments remain between some features seen in both instruments FOVs (as seen later on Fig.~\ref{fig:evolution_strands_sst}). 
This is due to the wideband images that are fundamentally different passbands at very different wavelengths and to uncompensated seeing variations over the FOV, which are also different for the red and the blue ends of the spectrum. However these misalignments (varying spatially and temporally) are corrected on the specific area of the FOV used for the analysis of coronal rain (Sect.~\ref{sec:physical_rain}).

\subsection{Co-alignment of the SST sequence with AIA}

\begin{figure}
	\resizebox{\hsize}{!}
	{\includegraphics[scale=0.8]{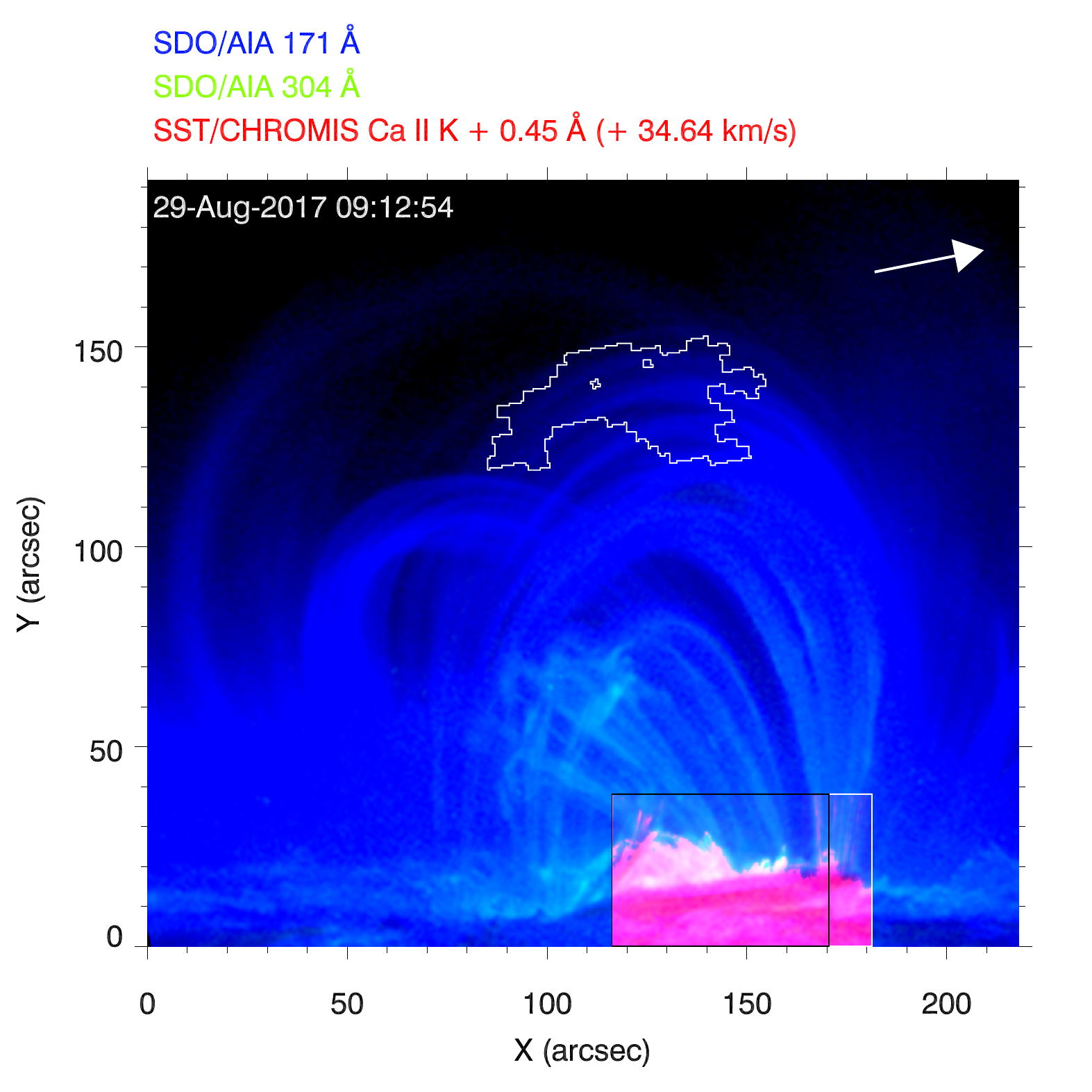}}
	\caption{RGB image combining two AIA channels: 171~\AA~in blue and 304~\AA~in green and one position in the red wing (+~0.45~\AA) of the \ion{Ca}{II}~K line scanned with CHROMIS in red. The white contour represent the regions detected with normalised power above $10\sigma$ (i.e. the same contour as the black one in Fig.~\ref{fig:img_aia_power_maps}). The black and white boxes represent, respectively, the FOV of the CRISP and CHROMIS instruments at the beginning of the SST observations. The white arrow indicates the north direction. The image is rotated compared to the one of Fig.~\ref{fig:img_aia_power_maps} for convenience. (An animation of this figure is available.)
    } 
	\label{fig:fov_aia_sst}
\end{figure}

As already stated in Sect.~\ref{sec:aia_data}, we used a different set of AIA images for the co-alignment procedure with the SST dataset than the main one used for the pulsations and thermal analysis. This was done for two main reasons: we needed the AIA high cadence (i.e. 12 seconds) to match as close as possible the one of the SST and the cleanest co-alignment possible between the channels. The latest matters especially because taking advantage of the high resolution of SST data, we aimed to look at small features. We further used some routines developed by Rob Rutten in order to carefully co-align the different channels of AIA.

For the co-alignment, we used common bright points seen in the SDO/HMI continuum and the CRISP wideband on the on-disk part of the CRISP FOV. For each time step, we computed the spatial shifts between the two filters with a cross-correlation technique. 

Coronal rain blobs are clearly visible in both chromospheric lines, for most of the dataset duration: in their red wing for the pulsating loop bundle and mainly at line core for the other loops of the SST FOVs.
A composite image made with the 171~\AA~and 304~\AA~AIA channels and one position in the red wing (+~0.45~\AA) of the \ion{Ca}{II}~K line (CHROMIS) is presented in Fig.~\ref{fig:fov_aia_sst}. 
The footpoint of the pulsating loop bundle is fully covered by CHROMIS but not by CRISP (FOV highlighted by the black box), due to its smaller extension on the x-axis (parallel to the limb in that case). 

A visual check in real time at the SST later on, around 11:50 (i.e. $\sim 36$ hours on Fig.~\ref{fig:light_curves_AIA}) in 
\halpha\ showed a significant decrease of rain in the same FOV. This seems to be confirmed by the 304~\AA~channel of AIA where almost no rain is seen in the bundle showing pulsations at that time (see Fig.~\ref{fig:rain_cycle} and Movie~1).

\subsection{Absolute intensity calibration for \halpha}
In the analysis presented in Sect.~\ref{sec:sst_analysis}, we estimated the rain blobs density using absolute intensities in \halpha. The procedure is as follows. Firstly, the intensity calibration to absolute units was performed by using the average line-profile at disk center (DC), taken one hour after the science data. A calibration was found by computing a factor that matches this profile to the Fourier Transform McMath-Pierce Fourier Spectrometer atlas \citep{brault_atlas, Neckel1999SoPh..184..421N} post convolution with the CRISP instrumental profile as in \cite{Jaime2013A&A...556A.115D}.
Secondly, due to the time difference between the rain dataset and the DC calibration data we applied an additional correction, based on the average intensity observed in the wideband channel (WB) from an on-disk cut-out of the rain dataset, converted to its DC counterpart using the Keith Pierce's limb-darkening model from \cite{Allen1973asqu.book.....A}. This correction is then the ratio of such DC counterpart and the WB DC from the calibration set.
We note that the cosine of the heliocentric angle $\mu$ of the on-disk region used is 0.05 and thus the time compensation suffers from whatever mismatch the limb darkening model has with regards to our data and the Sun in general. However, the latter was found to be a minor factor in the calibration, as it should be due to a short time-difference and the slow intensity variation of the mid-morning period.

\section{Thermal cycles analysis}\label{sec:cycles_analysis}

In order to have a full picture of the thermal evolution of this active region and especially the pulsating loop bundle, we combined:

\begin{itemize} 
    \item A DEM analysis and a time-lag analysis \citep[following the method introduced in][]{viall_evidence_2012}, presented in this section. This analysis is similar to the one conducted by \citet{froment_evidence_2015} for long-period intensity pulsations events observed on-disk with SDO/AIA.
    \item A full statistical analysis of the thermodynamics of the rain blobs observed with the SST, in a similar way that has been done in \citet{Antolin_Rouppe_2012ApJ...745..152A} and \citet{Antolin_2015ApJ...806...81A}. 
    Besides deriving the temperature, velocity, and width of these structures, we also computed their density and refine the temperature determination by estimating the non-thermal broadening of the chromospheric lines observed. This analysis is presented in Sect.~\ref{sec:sst_analysis}. 
\end{itemize}

\subsection{Evolution of the thermal structure with a DEM analysis}\label{sec:DEM}

\begin{figure}
	\centering
	\includegraphics[width=\hsize]{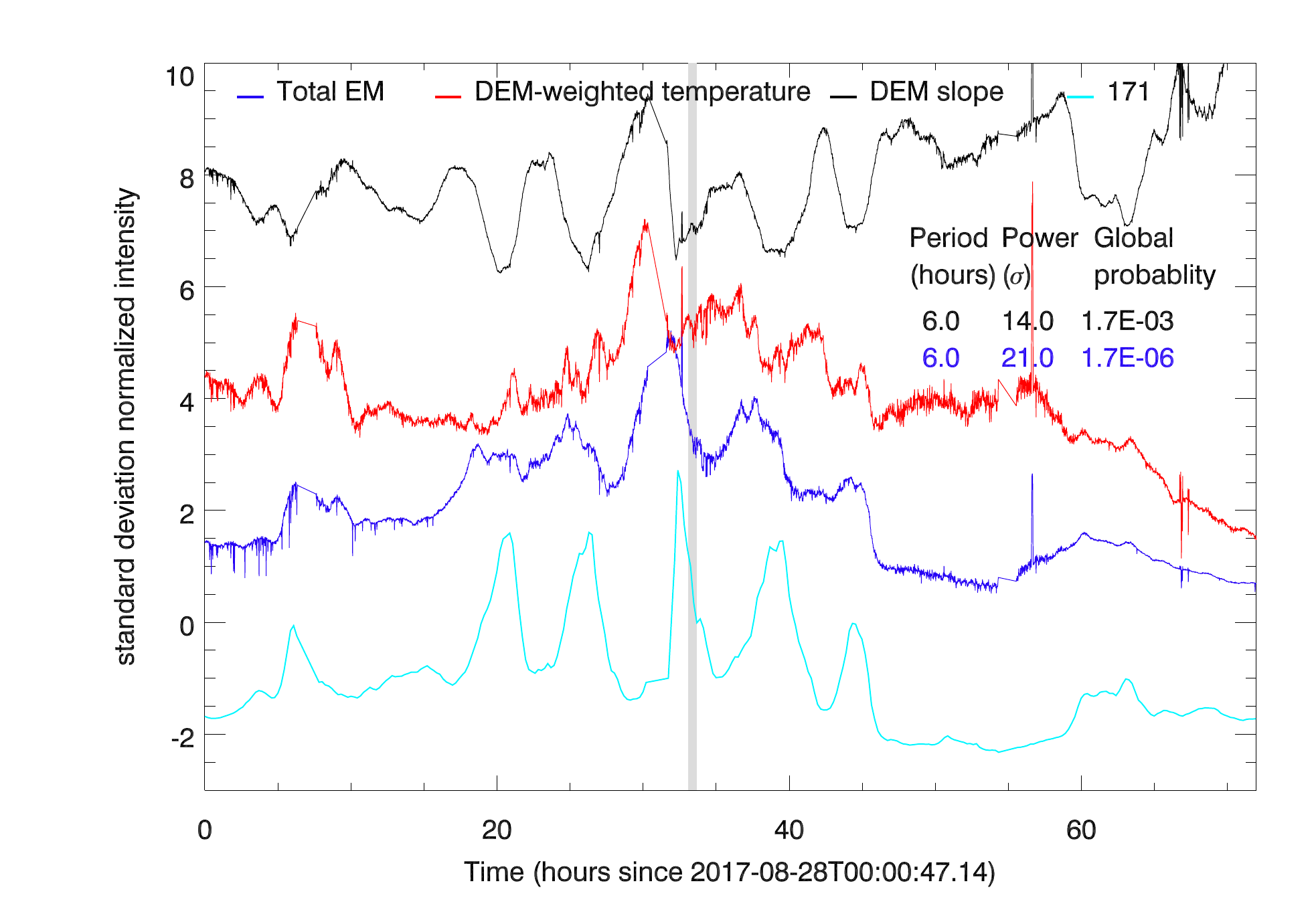}
	\caption{Evolution of three parameters extracted from the DEM analysis: DEM-weighted temperature, DEM slope and total Emission Measure (EM). These time series are averaged over the black contour displayed in Fig.~\ref{fig:img_aia_power_maps}. We included the 171 light curve from Fig.~\ref{fig:light_curves_AIA} for reference. All these quantities are normalised to the standard deviation and offset on the y-axis. If there is a detection, we display the corresponding period detected, the Fourier power and the associated random occurrence probability. The grey bar indicates the SST observations duration. (An animation of this figure is available.)}
	\label{fig:dem_curve}
\end{figure}

The DEM was computed using the method developed by \citet{cheung_thermal_2015}. As stated previously in Sect.~\ref{sec:aia_data}, we used here a cadence of 1~minute.
The DEM was computed for each pixel of the ROI and for the three days of the sequence (i.e. for about 4060 time steps).  We used the six coronal channels of AIA (335, 211, 193, 171, 131, and 94), allowing for a maximum time difference of $\pm1$ minute for the sextuplets grouped.

The DEM was computed in the range $\mathrm{log} \, T \in [5.5, 7.]$, using 16 temperature bins (i.e. a resolution of $\mathrm{log} \, T=0.1$). 
We did not push the DEM reconstruction under $\mathrm{log} \, T=5.5$ in order to avoid the known spurious temperature bump that can appear below this temperature and the accompanying overestimate of the DEM above $\mathrm{log} \, T=6.8$. This dependence of the DEM inversion results on the choice of the range of temperature explored is described in the Appendix B of \citet{cheung_thermal_2015}. 
We chose then to follow the evolution of three different parameters that we extracted from the DEM analysis:

\begin{itemize}
	\item the DEM-weighted temperature, that is, the averaged temperature for which each temperature bin is weighted by their emission measure. This measure is less sensitive to noise than the peak temperature (i.e. the temperature at the maximum of the DEM).
	\item a slope that we fit on the DEM distribution between $\mathrm{log} \, T\in [6.0, 6.3]$, in order to follow the evolution of the quantity of cool plasma captured by the DEM analysis. We focused only on the part of the cool wing where the EM is still high to ensure the best signal-to-noise ratio possible.
	\item the total emission measure (integral of the DEM over the electron temperature), proportional to the squared electron density along the LOS.
\end{itemize}

In Fig.~\ref{fig:dem_curve}, we display the corresponding time series for three quantities, averaged over the 10~$\sigma$ contour. 
On top of doing an analysis of the Fourier spectra, we compute the cross-correlations between these parameters, exploring time shifts of $\pm 180$~minutes.

The time series of the DEM slope and the total emission measure are showing clear pulsations. For both parameters, we find significant Fourier power at 6.0 hours (46.1~$\micro$Hz) at, respectively, 14$\sigma$ and 21$\sigma$ (i.e. $1.7 \times 10^{-3}$  and $1.7 \times 10^{-6}$ of random occurrence probability, derived with the same method as used in Sec.~\ref{sec:period_analysis}). We can indeed easily recognise the five central pulses that were also present in the intensity curves. 
However, we can immediately notice that the evolution of the temperature is particularly noisy. No clear periodicity stands out from the Fourier analysis. It makes the analysis of the time delays with the other DEM parameters particularly difficult since the shapes of individual curves are too different. We therefore decided to limit the time-lag analysis between 27 hours and 45 hours after the beginning of the sequence, where pulses seem to be visually present in the temperature time series too. We find a time lag of 1.1 hour between the temperature and the total emission measure curves and that the temperature and slope curves are in phase. The corresponding peak cross-correlation values are, respectively, 0.68 and 0.53.

The variations in amplitude are comparable with previous results of \citet{froment_evidence_2015}. Indeed, the variations in temperature are relatively small but there are large variations in total emission measure and slope. The DEM-weighted temperature varies between 2~MK and 2.5~MK (relative amplitude of $20\%$) during the five central pulse time period while the total emission measure varies between $6.3 \times 10^{26} \; \mathrm{cm}^{-5}$ and $15.7 \times 10^{26} \; \mathrm{cm}^{-5}$ (relative amplitude of $60\%$). On the same time period, the DEM slope varies between 3.4 to 7.2 (relative amplitude of $53\%$).  In \citet{cheung_thermal_2015}, the error on the total EM is estimated be less than $10\%-20\%$ and 0.2 $\mathrm{log} \, T$ for the temperature. From the DEM slope, we estimated an error of less than 1 for the five central pulses, with variations not affecting their periodicity.

In conclusion, while the quantity of cool plasma decreases along the LOS (when the value of the DEM slope increases), the temperature increases, followed by an increase in density approximately 1~hour later. On top of that, the evolution of the quantity of cool plasma along the LOS and of the density is periodic (along with an indication that the temperature evolution is periodic as well). Such as for the pulsating loops event of \citet{froment_evidence_2015}, we thus observe a periodic change of the thermal structure of these loops, from a multi-thermal distribution to a more isothermal distribution, with an indication that the density evolution is delayed of about 18$\%$ of the cycles period compared to the temperature evolution, giving an estimate if the evaporation timescale. This is already a clear evidence that these loops undergo TNE cycles.

\subsection{Cooling signatures from time-lag maps}\label{sec:time_lag}

\begin{figure*}
	\resizebox{\hsize}{!}
	{\includegraphics[trim={0cm 6cm 5cm 0cm},clip]{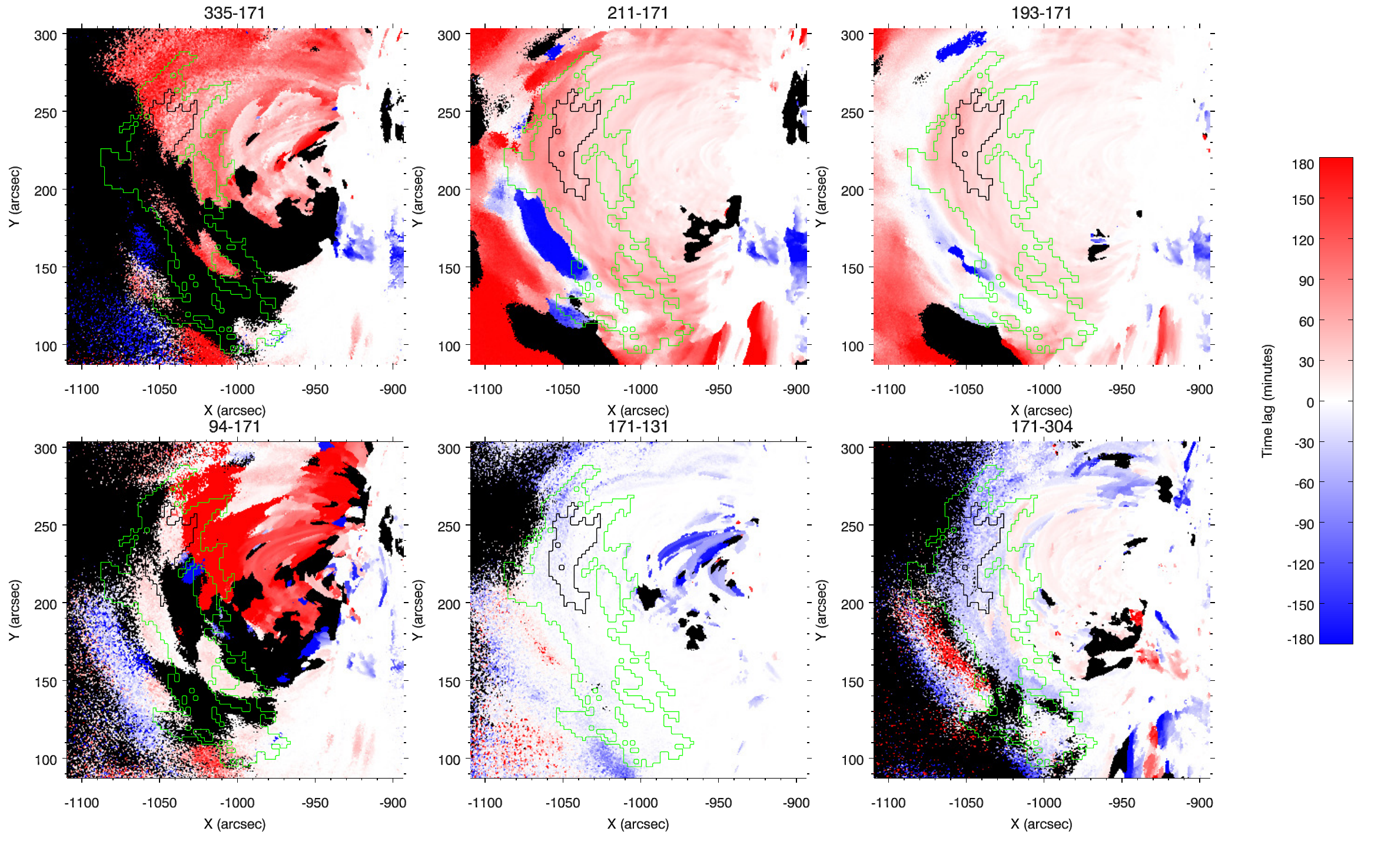}}
	\caption{Time-lag maps for the same FOV as in Fig.~\ref{fig:img_aia_power_maps}. We selected six pairs of AIA channels: 335-171, 211-171, 193-171, 94-171, 171-131, and 171-304. Black areas represent peak cross-correlations values under 0.2. The green and black contour are, respectively, the $5\sigma$ (extended contour) and $10\sigma$ contours of detection.
	}
	\label{fig:time_lag_maps}
\end{figure*}

\begin{table*}
	
	\caption{Median time lags from the cross-correlation technique (computed between 15 and 45 hours after the beginning of the sequence) and taken over the $10\sigma$ pulsations detection contour of Fig.~\ref{fig:time_lag_maps}.}
	\label{table:time_lag}
	\centering
	\begin{tabular}{c c c c c c c}
		\hline\hline
		Pairs of channels & 335-171 & 211-171 & 193-171 & 94-171 & 171-131 & 171-304   \\
		Cross-correlation time lag (minutes) & 88 & 52 & 31 & 18  & -4 & -34 \\
		\hline
	\end{tabular}

\end{table*}

The global cooling properties of this active region are studied by mapping the time lags between selected pairs of AIA channels.
This technique, first introduced in \citet{viall_evidence_2012}, is based on the computation of cross-correlations between two EUV light curves produced in different channels for the same spatial location. Reiterated for all the pixels of a FOV, we can eventually obtain the map of time lags between two channels by selecting the peak cross-correlation values. Although AIA channels have broad temperature response functions, showing strong secondary peaks for some of them, it is possible to get a global idea of how the bulk of the active region plasma is cooling, the intensity gradually peaking up from one channel to another according to their peak temperature response \citep[see e.g. Fig. 5 of][]{auchere_coronal_2018}. This is made possible since the coronal emission is dominated by the cooling phases. During the heating phases the density of the plasma is smaller than during the cooling phases, making the temperature rise imperceptible. This property has been observed by many authors \citep[e.g.][]{warren_hydrodynamic_2002,ugarte-urra_investigation_2006, ugarte-urra_active_2009,viall_patterns_2011,viall_survey_2017} and can be both explained so far by uniformly distributed heating models \citep[e.g][]{viall_modeling_2013,bradshaw_patterns_2016} as well as TNE models \citep[e.g.][]{lionello_can_2016, Winebarger_2016ApJ...831..172W, froment_long-period_2017}. This is also clearly the case for the observations we analyse in this paper (see Sect.~\ref{sec:DEM} and Fig.~\ref{fig:dem_curve}).
 
We explored time shifts from $-180$ to 180 minutes (i.e. about half the period of the pulsations detected), only between 15 hours and 45 hours after the beginning of the sequence, that is, the central 30 hours of the five central pulses. We thus keept only the times when the loops are well above the limb to minimise the displacement of the structures due to the solar rotation. We also did the same analysis dividing this central window in 12 hours time windows in order to check the influence of the active region evolution and displacement on our results. With these shorter time windows, we found similar time lags as the one presented below.

Fig.~\ref{fig:time_lag_maps} shows the time-lag maps for six selected pairs of channels: 335-171, 211-171, 193-171, 94-171, 171-131, and 171-304. We explored pairs of channels that include 171, where the pulsating signal is the strongest and because we can then directly compare our findings with the ones of \citet{auchere_coronal_2018}. Note that 171 is the first channel of the last two explored pairs, and it is the opposite for the first four pairs. We took the convention of a coronal plasma that would cool from at 2.5~MK (the peak temperature of 335 in the corona) to at least 0.08~MK (the peak temperature of 304), and thus place 171 (peak temperature at 0.9~MK) in the middle of the channel sequence.

The median time lags over the $10\sigma$ Fourier detection contour are displayed in Table~\ref{table:time_lag}. We chose to show the median over the average since the former is not sensitive to extreme values (that can be found in the contour for example in the 94-171 map). 

The 335-171, 211-171, 193-171, and 94-171 maps are mostly showing positive time lags. In the contour of detection the median time lags are, respectively, 88, 52, 31, and 18 minutes for these four pairs of channels. It means that the plasma is mostly observed in a state of cooling, from at least the peak temperature response of the 335 channel (2.5~MK) toward the peak  temperature response of the 171 (0.9~MK). In between, the intensity peaks first in the 211 channel (2~MK), 193 (1.5~MK), and then 94 (1~MK). For this latest channel, we note that in the core of the active region the time lags are much higher (above 120 minutes) revealing that in this region the plasma cools down from the other temperature peak of 94 (7~MK). This feature is already known from other studies \citep{viall_evidence_2012, froment_evidence_2015, viall_survey_2017}.
For most of these maps and as already seen in \citet{froment_evidence_2015}, it seems that the pulsating loops do not have a different pattern of time lags compared to the loops that are not showing periodic pulsations. 
Very high time lags excursions within the pulsating loop bundle seem to be present but should not be over interpreted since, as seen in  Sect.~\ref{sec:aia_data}, the LOS in the 94 channel seems to be dominated by the plasma of other structures than the pulsating loops.

In the 335-171, 211-171, and 193-171 maps, it seems that there are longer time lags toward the outer edge of the loop bundle. This is consistent with the fact that for similar heating conditions (that we assumed for these loops bundles whose footpoints are close to one another) longer loops will have longer cooling time.

However, for two different active regions with TNE cycles, we can reach a quite different conclusion. The time lags values found here for the 335-171, 211-171, 193-171, and 94-171 pairs are generally larger than the ones found in \citet[][e.g. 211-171, about 10~min compared to 52~min here]{auchere_coronal_2018}. Yet, the loop bundle analysed in their paper was significantly longer compared to the one we analyse here (2 to 3 times longer considering the position of the footpoints and a semi-circular loop approximation). Albeit their large difference in loop length, both pulsating events have a similar cycle period (about 6~hours). In order to cool down the plasma of a longer loop, assuming a similar density and temperature range, and maintaining the same cycle period, the cooling time has thus to be faster. We can speculate that to achieve such fast cycles for a loop bundle as long as the one presented in \citet{auchere_coronal_2018}, the heating strength was probably very different than for the active region presently studied. This is consistent with the simulations of \citet{froment_occurrence_2018}.

The 171-131 map is dominated by negative time lags in the core of the active region and zero time lags for the external loops (median time lag of $-4$~min in the detected contour and even lower along the legs of the bundle). This is, thus, very similar to the cases observed in \citet{froment_evidence_2015}. Indeed, zero time lags dominate in the pulsating loops area while negative time lags appear when the high temperature peak of 131 (11~MK) is picked up by the cross-correlation method. In \citet{froment_evidence_2015} , as in for example, \citet{viall_survey_2017}, the interpretation of these zero or close to zero time lags was that the plasma temperature of the bulk of the active region was not decreasing below the peak temperature of 171 at 0.9~MK, and thus not reaching the cool peak temperature of 131 at 0.5~MK. However, here we observe coronal rain in the 304 channel throughout the sequence studied. Since the plasma is cooling under 0.5~MK, we would therefore expect to detect positive time lags in these areas for the 171-131 pair.

On the other hand, the 171-304 map displays a bimodal behaviour. While the very apex of the pulsating loop bundle is dominated by negative time lags (median time lag of $-34$~min in the detected contour), the internal part of the active region (apex of the internal loops and legs of the pulsating bundle, and core of the active region) is dominated by positive time lags of few tens of minutes.
The 304 channel has two main components with one peak at 0.08~MK (absolute maximum in the temperature response function) and a secondary peak at $\sim1.8$~MK. It thus seem that the apex region of the pulsation is dominated by 1-MK plasma, with the temporal succession 211, 304, 193, 94, and then 171 (see Table~\ref{table:time_lag}), while the plasma in the pulsating loops legs is dominated by the transition-region rain signature.

Reexamining now the 171-131 map, we can better understand why the intensities in the 171 and 131 channels are almost in phase around the pulsating loops apex captured by the black contour. If some of the rain appear to originate from this region (see Fig.~\ref{fig:rain_cycle} and Movie~1), most of the rain originates from the apex of the internal loops of the bundle and the LOS seem dominated by coronal plasma at $\sim1$~MK. However, the presence of zero time lags at the apex of the internal loops and legs of the pulsating bundle remains a puzzle. This plasma response may be affected by the emission from the foreground and background of the pulsating loop bundle. We could reduce this effect by computing the phase difference between the main Fourier components for each channel pairs, as introduced in \citet{froment_evidence_2015} and applied in \citet{auchere_coronal_2018}. However, the use of this method would require a significant power in the pulsating loop legs, which is not the case. 

\subsection{On the use of time-lag analysis with the 171-131 channel pair}

As we developed in the previous section, the domination of zero time lags in the 171-131 maps has been generally interpreted as evidence that the plasma was re-heated before having the time to cool down to the 0.5~MK peak response of the 131 channel. This feature is very commonly observed. In \citet{viall_transition_2015, viall_survey_2017} for example, the 171-131 maps show mostly zero time lags, sometimes negative time lags, but almost never positive time lags.
On the contrary, in \citet{auchere_coronal_2018} that studied a region with a large coronal rain production seen in the 304 channel, the 171-131 lags were  positive. In this paper, the authors demonstrated that the LOS was dominated by the background and foreground emission of the raining loops, the 171-131 time lag increasing from 4 to 15 minutes after isolating the contribution of the loops in TNE (via the main Fourier component).
Here, we witness TNE cycles with coronal rain, yet zero and negative time lags dominate the 171-131 map. It is unclear why positive time lags do not show up in the rain-dominated regions as here isolating the contribution of the loops in TNE is not meaningful everywhere along their length.

In \citet{viall_evidence_2012} two other explanations were mentioned to explain the zero time lags for this channel pair: a very rapid cooling from 0.9~MK to 0.5~MK, that is, the respective temperature response peaks of 171 and 131, or that the contribution near 0.7~MK is underestimated in the 131 response function \citep{testa_testing_2012}\footnote{Since then, the AIA response functions is better known and it seems that the CHIANTI v.9 \citep{dere_chiantiatomic_2019} is better taking in account these contributions.}.
For our present observations, a non-resolvable cooling from 0.9~MK to 0.5~MK may be the answer.
Even though understanding in detail why 171-131 time-lag maps are showing preferentially zero time lags is beyond the scope of this paper, we can conclude that the absence of time lag between 171 and 131 alone is generally not enough to state that the coronal plasma does not cool down through the temperature peak response of 131. Interpreting this result is of course far from being trivial. However, this particular point is very important since it is an argument used to test nanoflare scenarios \citep[][]{bradshaw_cooling_2010, bradshaw_patterns_2016}. It has also been used to differentiate incomplete-condensation TNE cycles \citep{froment_evidence_2015} from complete-condensation TNE cycles \citep{auchere_coronal_2018}. In the incomplete-condensation scenario, the plasma is believed to never reach transition-region temperatures \citep{mikic_importance_2013}.  

We suggest, therefore, that the use of the 171-131 time lag information needs to be revisited and that 304 should be added in time-lag analyses. This is generally not the case even if the time-lag map method is now widely used \citep{viall_survey_2017,Winebarger_2018ApJ...865..111W, barnes_understanding_2019}.

\begin{figure*}
	\centering
		\resizebox{\hsize}{!}
	{$\begin{array}{cc}
		\includegraphics[trim={0cm 0cm 8cm 0cm},clip]{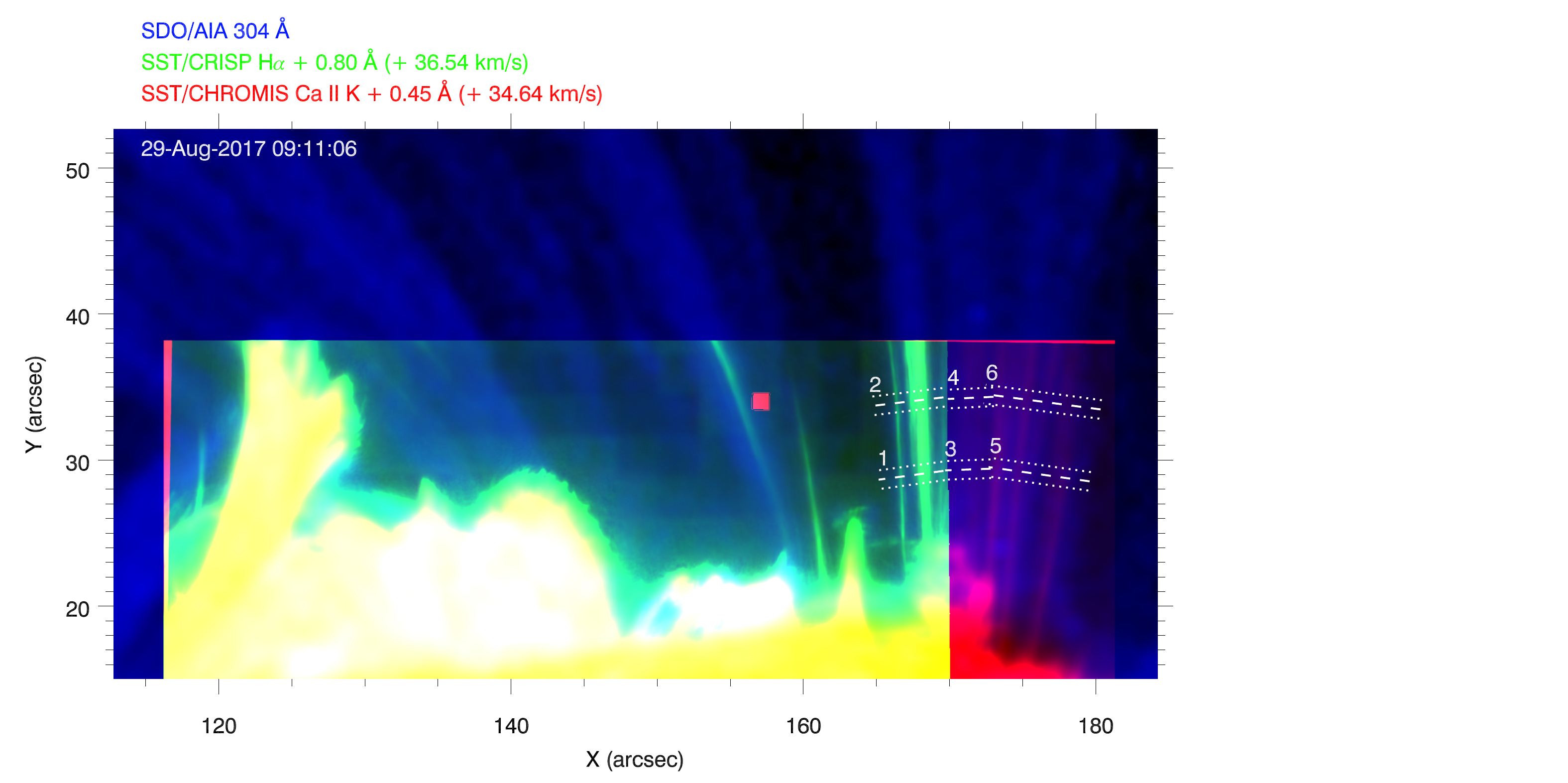} &
		\includegraphics[trim={0cm 0cm 8cm 0cm},clip]{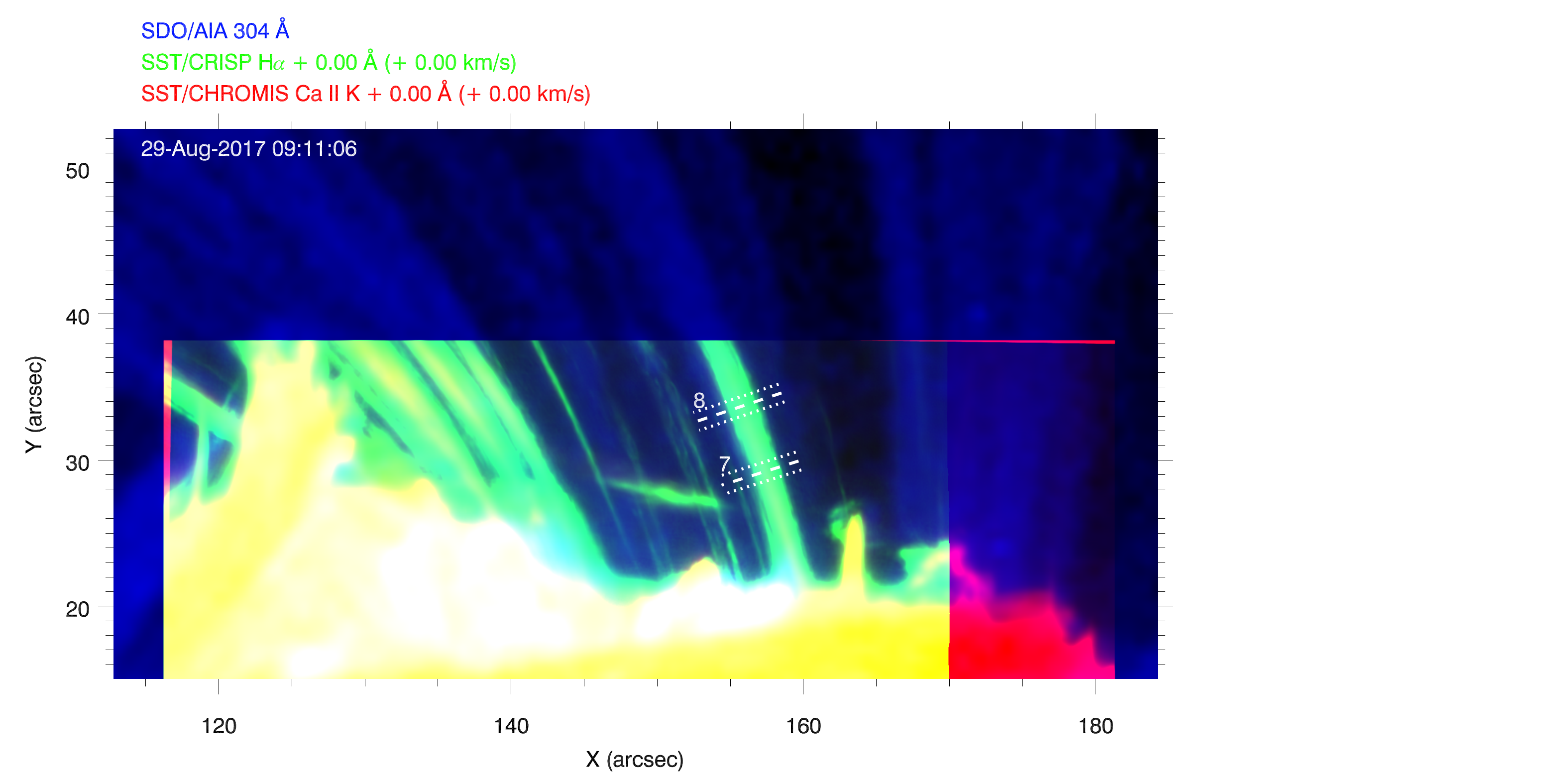}
	\end{array}$
	}
	\caption{Zoom on the the coronal rain in the SST FOV. Images are stacked from 09:07:06 to 09:11:06 in order to highlight the display of the different bundles. The dashed lines show the different paths chosen for the analysis and the parallel dotted lines show the thickness covered by the slices constructed around these paths. 
	Left: RGB image combining the AIA 304~\AA\ in blue, red wing \halpha\ in green, and red wing \ion{Ca}{II}~K in red.
	Right: Same FOV with \halpha\ line center in green and \ion{Ca}{II}~K line center in red.
	}
	\label{fig:sst_path}

\end{figure*}

\begin{figure*}
	\resizebox{\hsize}{!}
	{\includegraphics[trim={0cm 0cm 8cm 0cm},clip]{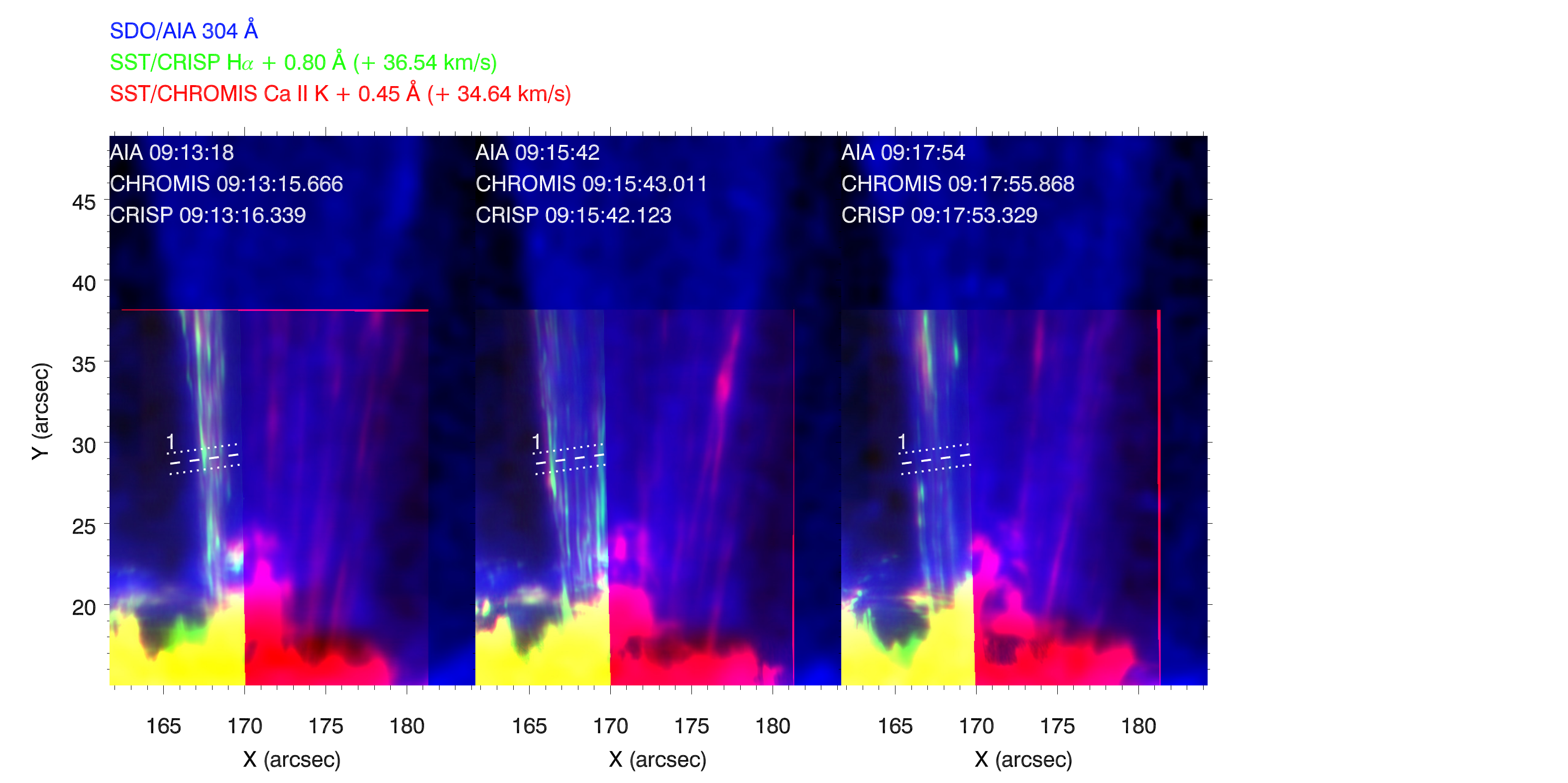}}
	\caption{Close-up on the northern leg of the loop bundle showing long-period EUV pulsations. RGB images at different steps of the SST sequence. For each step we show a combination of the AIA 304~\AA~channel in blue, one position in the red wing (+~0.80~\AA) of the \halpha\ line in green, and one position in the red wing (+~0.45~\AA) of the \ion{Ca}{II}~K line in red. Path 1 is displayed as a context for Fig.~\ref{fig:strands_sst}. (An animation of this figure is available.)}
	\label{fig:evolution_strands_sst}
	
\end{figure*}	
\begin{figure*}
	\resizebox{\hsize}{!}
	{\includegraphics[trim={0cm 0cm 0cm 4cm},clip]{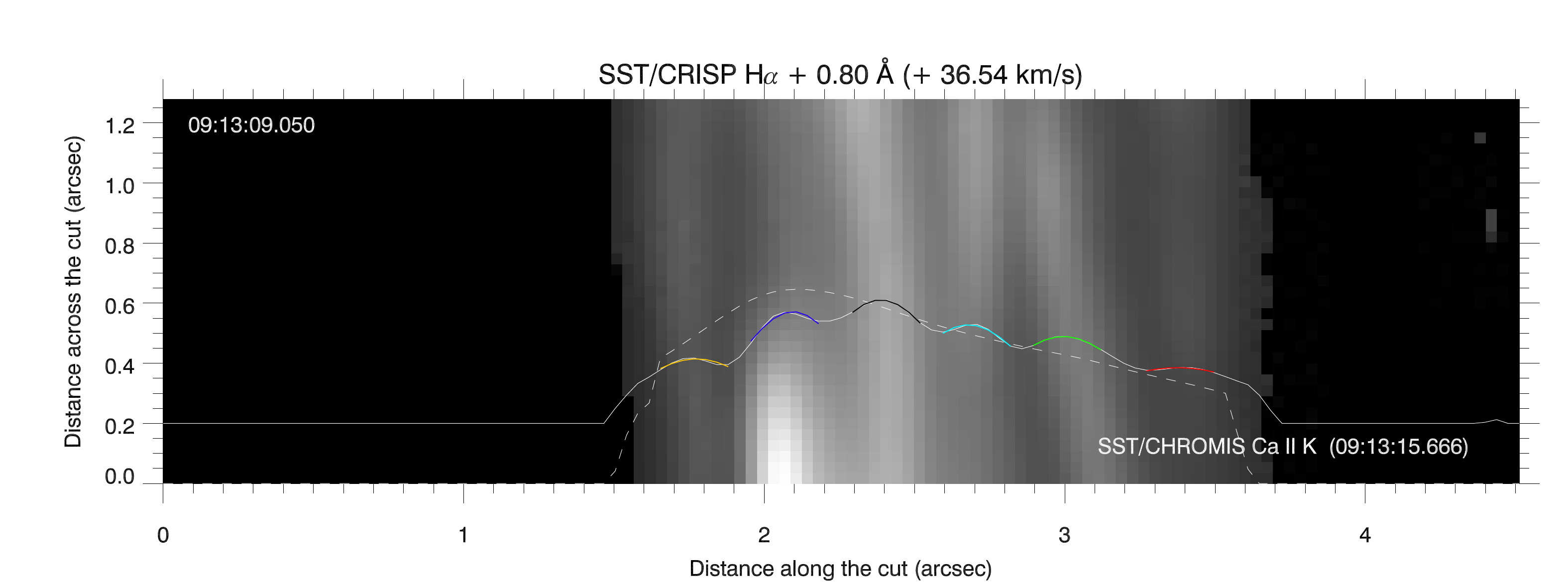}}
	\caption{\halpha\ slice corresponding to path 1 at the same time and wavelength position as the right-most panel of Fig.~\ref{fig:evolution_strands_sst}. Six rain strands are clearly visible with widths of about $0.2^{\prime\prime}$ (i.e. 140 km). The black area corresponds to locations where no rain blobs are detected (under the intensity threshold). The white solid line corresponds to the average intensity (arbitrary units) along the path for the slice. The different coloured lines are the widths of the spatial Gaussian fits to the different blobs. The white dashed lines corresponds to the average intensity for the same slice at the closest time and Doppler shift in \ion{Ca}{II}~K line.}
	\label{fig:strands_sst}
\end{figure*}

\section{Full catastrophic cooling to chromospheric temperatures}\label{sec:sst_analysis}

After a general overview of the thermal evolution of the active region and in particular of the pulsating loop bundle, we now focus on the catastrophic cooling of the plasma during the third cycle detected, namely, one rain cycle.

\subsection{Morphology and temporal evolution of the rain}

The SST observations occur when the minimum temperature is reached at the apex of the loop bundle (as seen from Figs.~\ref{fig:light_curves_AIA},~\ref{fig:rain_cycle}, and ~\ref{fig:dem_curve}) and when the evacuation phase is still ongoing (the total EM has not stopped decreasing), and during the decreasing phase in the AIA intensities in the $10\sigma$ contour. This is similar to what was described in \citet[][Fig.~9 at $t_2$]{froment_long-period_2017}.
Moreover, the absence of rain about 36~hours after the beginning of the AIA sequence, pointed out in Sect.~\ref{sec:sst_data}, is clearly simultaneous to the evaporation-dominated phase of the fourth cycle (Fig.~\ref{fig:dem_curve}), that is, when the EM and thus the density is increasing and no rain is expected at this point of the cycle.

Figure~\ref{fig:sst_path} displays a zoom on the coronal rain observed near the footpoints of the pulsating loops. The first panel displays a stack of a few time steps in order to highlight the rain paths in the red wing of the \halpha\ 
and \ion{Ca}{II}~K lines since these are the line positions where we observe rain in the pulsating loop. 
The rain has maximum intensity at a Doppler offset of about 37~\kms.
This loop footpoint lies behind the limb, the broad range of LOS velocities in which rain is present (i.e. 20 to 60~\kms) revealing the stretch of the loop bundle on the LOS (i.e. the different inclination angles of the loop strands compared to the plane of the sky (POS)).
The other bundles located in the core of the active region, although showing some rain blobs in the red wing of the two lines, mostly show rain with LOS velocity close to 0~\kms\ as these loops mostly lie in POS.

Figure~\ref{fig:evolution_strands_sst} is showing different time steps in the SST sequence. At the beginning of the sequence, one main sub-bundle is seen mostly in \ion{Ca}{II}~K. From 09:11~UT, three distinctive sub-bundles are visible in \ion{Ca}{II}~K.
We checked the rain morphology and evolution for different pairs of Doppler shifts in \halpha\ and \ion{Ca}{II}~K.
The rain dynamics and morphology seem similar in both lines.
The main differences come from the fact that the \ion{Ca}{II}~K observations suffer more from seeing effects than the \halpha\ ones. The Movie~3 associated with this figure shows that coronal rain blobs are falling down  more or less continuously toward the loop footpoint in characteristic clumps.
From the rain crossing path~1, perpendicular to the trajectory (detection  method later presented in Sect.~\ref{sec:results_rain}), most of the time at least one blob was detected per time step with a maximum of 7 blobs detected for one time step. We note a lack of rain at the first 2 min of the sequence, only because of the time it takes for the first blobs to reach path~1 (see Movie~3). In the other rain strand, analysed in Sect.~\ref{sec:results_rain} (see the right panel of Fig.~\ref{fig:sst_path}), most of the rain detection occurs in the first 10~min of the sequence, denoting the difference in dynamics for these rain events.  

In Fig.~\ref{fig:strands_sst}, we show one time step of the rain detection within the slice produced around path~1. It reveals the strand-like structure that we were able to capture with the high-resolution observations in \halpha. The approximated width of these stands appears here to be 0\farcs2.  
Unfortunately, the seeing being generally worst at the blue end of the spectrum (like for this time step for example), these fine structures are not always well  captured in \ion{Ca}{II}~K.  

\subsection{Velocity, temperature, and density measurements of rain blobs}\label{sec:physical_rain}

	\subsubsection{Method}\label{sec:method_rain}
	In order to measure the kinetic and thermodynamic properties of the rain in \halpha\ and \ion{Ca}{II}~K, we developed an automatic routine that scans and detects all the rain blobs crossing a given path. We took various perpendicular paths to the local trajectory of the rain and applied the scanning routine to measure LOS (Doppler) and POS velocities, temperatures, non-thermal line widths, and densities.
	
	This routine builds on previous versions that use the same or similar techniques to measure these physical quantities \citep{Antolin_etal_2012SoPh..280..457A,Antolin_2015ApJ...806...81A}. On top of including densities and multi-spectral temperature measurements here, the version of the routine used benefited from several improvements that are detailed throughout this section.
	In particular, since the velocity of the blobs is significant and the cadence of the observation is not high enough to ensure that between two consecutive time steps every blob crosses the chosen path, the routine works by first taking parallel cuts above and below the path. This defines a FOV large enough to capture most rain blobs crossing the path at least once between two consecutive time steps. The different paths were placed nearly perpendicularly to the rainy loop bundles, at different heights. We carefully checked that no other structures like small prominences or surges were crossing the slices throughout the time sequence. 
	
\begin{figure*}
	\resizebox{\hsize}{!}
	{\includegraphics[trim={0cm 0cm 0cm 2cm},clip]{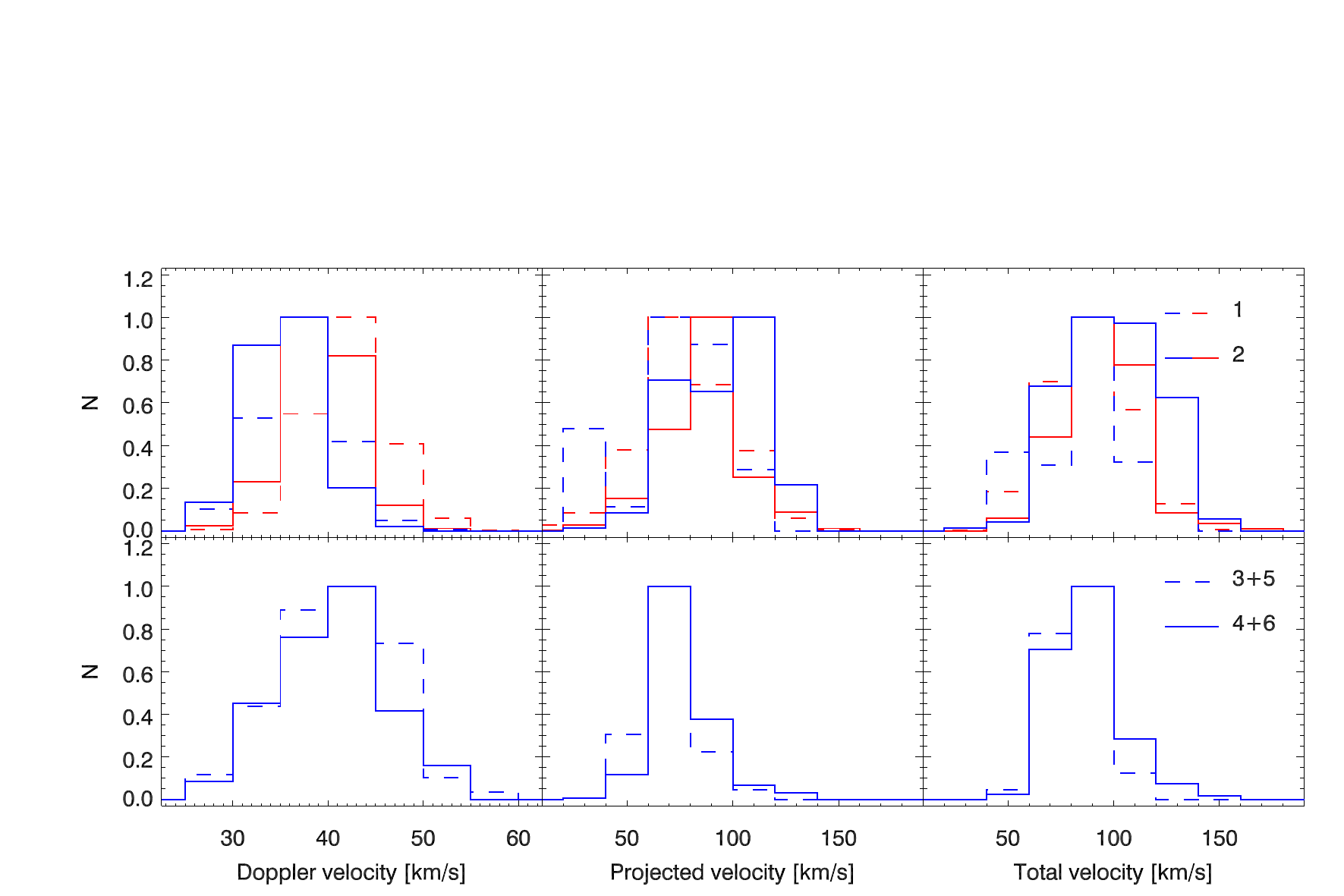}}
	\caption{Normalised histograms of the velocities for the rain in the pulsating loop bundle. From left to right: Doppler velocity, projected velocity, and total velocity.  We only show the measurements that have an error below 15\% of the measured value. Top row: measurements for paths 1 and 2, in red for \halpha\ and in blue for \ion{Ca}{II}~K. Bottom row: measurements for paths only covered by CHROMIS (\ion{Ca}{II}~K). Paths 3 and 5, and 4 and 6 are combined given the small number of rain pixels in paths 3 and 4.}
	\label{fig:stats_vel}
\end{figure*}

	For each time step the routine then selects all pixels within the defined FOV with large enough signal-to-noise ratio (i.e. the rain pixels) and performs a fit of the spectra over wavelength for each pixel. These fits are used for the temperature and Doppler velocity measurements.
	Due to the combination of the relatively low cadence used (about 7 and 14 seconds for, respectively, CRISP and CHROMIS), the high resolution of these instruments and the high POS velocities of the rain (between $\sim 50~\mathrm{km} \, \mathrm{s}^{-1}$ to $\sim 150~\mathrm{km} \, \mathrm{s}^{-1}$, see Fig.~\ref{fig:stats_vel}), rain spectra can come from multiple blobs passing through one single pixel. Taking the example of two blobs moving at a POS velocity of $\sim 100~\mathrm{km} \, \mathrm{s}^{-1}$ one after the other through one CRISP pixel, we expect to see the two blobs imprint in the spectra if they have at least a difference in Doppler velocities of $\sim 9~\mathrm{km} \, \mathrm{s}^{-1}$. For smaller differences in Doppler velocities, only one blob should be visible in the spectra since before a second blob arrives, the wavelength position corresponding to the blob Doppler shift would have already been scanned.
	Since the routine we used is meant to treat thousands of spectra automatically it is very difficult to tailor the fitting procedure for various cases of multiple Doppler shift of rain blobs components. For that reason, we chose to only keep the spectra that are as close as possible to a Gaussian, that is, a single blob and hence use a single Gaussian fit with multiple validation criteria (relative position and intensity of the Gaussian and main profile peak).

	\paragraph{Velocity.} The Doppler velocity is calculated as in \citet{Antolin_etal_2012SoPh..280..457A}. For a specific rain pixel the average Doppler velocity is obtained with three different methods, and the error in the calculation is the standard deviation between the three different measurements. These are, respectively, the first moment with respect to wavelength, the value corresponding to the maximum of the spectral line profile, and the value corresponding to the maximum of the Gaussian fit \citep[see][for further details]{Antolin_etal_2012SoPh..280..457A}.
	
	For a given path, the POS velocities for blobs crossing the path were calculated as follows. We first took perpendicular cuts at each point along the given path and we made time distance diagrams for each cut over a short time interval over which blobs cross the path. Since the given path was chosen to be locally perpendicular to the rain trajectory, any blob will on average move along one of the perpendicular cuts to the path, thereby producing a distinctive intensity feature in the respective time-distance plot for this cut. The slope of the intensity feature in the time-distance plot gives the POS velocity of the blob. We measured this slope through intensity contours and also by tracking the intensity maxima. An estimate of the error in each POS measurement is given by the standard deviation between both methods, which includes also setting different levels of intensity threshold for the contours (we choose 60\% and 30\% of the intensity maxima within the time-distance plot). 
	
	The total velocities were then computed as  $\sqrt{\nu_{\mathrm{Doppler}}^2+\nu_{\mathrm{POS}}^2}$.
	
	\paragraph{Temperature and non-thermal line broadening.} As in \citet{Antolin_Rouppe_2012ApJ...745..152A}, an estimate of the temperature of the plasma emitting in \halpha\ is extracted from:

	\begin{equation}\label{eq:Halpha_temp}
	    \sigma_{\mathrm{H}}=\frac{\lambda_{0,\mathrm{H}}}{c}\sqrt{\frac{k_{\mathrm{B}}T_{\mathrm{H}}}{m_{\mathrm{H}}}+\xi_\mathrm{H}^2},
	\end{equation}
	
	where $m_{\mathrm{H}}$ is the Hydrogen mass, $\sigma_{\mathrm{H}}$ is the width of the Gaussian fit of the \halpha\ line profile, $\lambda_{0,\mathrm{H}}$ is the centre wavelength of the profile at rest, $c$ is the speed of light, $k_{\mathrm{B}}$ is Boltzmann's constant and $\xi_\mathrm{H}$ is the non-thermal velocity seen in \halpha. 
	This non-thermal broadening term represents  only the broadening of the profile from micro-turbulence since we assumed the instrumental broadening to be negligible. The extracted value for the temperature therefore corresponds to an upper limit. This is even further the case if we also assume a negligible non-thermal line broadening from micro-turbulence. Indeed, in \halpha, the thermal broadening is likely to dominate over non-thermal effects because of the low atomic mass of Hydrogen, \citep[see e.g.][Section 4.1]{cauzzi_solar_2009}.
	
	In the present paper, we also have simultaneous measurements in the \ion{Ca}{II}~K line and thus we can express similarly the line broadening of the plasma emitting in \ion{Ca}{II}~K:
	\begin{equation}
	    \sigma_{\mathrm{Ca}}=\frac{\lambda_{0,\mathrm{Ca}}}{c}\sqrt{\frac{k_{\mathrm{B}}T_{\mathrm{Ca}}}{m_{\mathrm{Ca}}}+\xi_\mathrm{Ca}^2},
	\end{equation}
	where the quantities are similar as for Equation~\ref{eq:Halpha_temp} but for the \ion{Ca}{II}~K line.

	While the relative high atomic mass of Calcium (factor of 40 compared to Hydrogen) does not allow us to extract directly an estimate of the temperature, we can take advantage of this mass difference to compute temperatures by combining \halpha\ and \ion{Ca}{II}~K measurements. We thus achieve more precision by estimating the non-thermal broadening term, as in \citet{Ahn_2014SoPh..289.4117A}, for example. 
	
	For that we assumed that the temperatures of the formation of these lines are the same, $T=T_{\mathrm{Ca}}=T_{\mathrm{H}}$, and that the non-thermal line broadening term is the same, $\xi=\xi_\mathrm{H}=\xi_\mathrm{Ca}$. This hypothesis is supported by the similar morphology and velocity of the rain we observed in both lines. Then we can estimate both quantities as:
	
	\begin{equation}\label{eq:T}
    	T = \frac{m_{\mathrm{H}}*m_{\mathrm{Ca}}}{m_{\mathrm{Ca}}-m_{\mathrm{H}}}\frac{c^2}{k_B}\Bigg[\left(\frac{\sigma_{\mathrm{H}}}{\lambda_{0,\mathrm{H}}}\right)^2-\left(\frac{\sigma_{\mathrm{Ca}}}{\lambda_{0,\mathrm{Ca}}}\right)^2\Bigg],
	\end{equation}
	
	\begin{equation}\label{eq:nth}
    	\xi = c\sqrt{\frac{m_{\mathrm{Ca}}*\left(\frac{\sigma_{\mathrm{Ca}}}{\lambda_{0,\mathrm{Ca}}}\right)^2-m_{\mathrm{H}}*\left(\frac{\sigma_{\mathrm{H}}}{\lambda_{0,\mathrm{H}}}\right)^2}{m_{\mathrm{Ca}}-m_{\mathrm{H}}}}.
	\end{equation}
	
	Both Equation~\ref{eq:T} and \ref{eq:nth} are true only for the condition $\frac{\sigma_{\mathrm{H}}}{\lambda_{0,\mathrm{H}}}>\frac{\sigma_{\mathrm{Ca}}}{\lambda_{0,\mathrm{Ca}}}$.
	
	For this method, it is critical that we analysed the same element of plasma while combining both lines. Given the misalignments that remain after the co-alignment procedure (see Sect.~\ref{sec:sst_data}), we re-coaligned semi-manually the scans in the different slices. Even though this procedure suffers from the differences in resolution for the strands visible in CRISP and CHROMIS, the seeing being worse in the blue part of the spectrum (see Fig.~\ref{fig:strands_sst}), we found that this extra alignment step has, in fact, minimal effect on the final distribution of temperatures.
	
	\paragraph{Density.} The densities are determined from the emission measure in \halpha, defined as $EM_{\mathrm{H}} = \int_0^{D}n_{e}^{2}dz$, where $n_e$ is the electron density along a given LOS through a blob, and the length $D$ is the region along the LOS emitting in \halpha. Indeed, \citet{Gouttebroze_etal_1993AAS...99..513G} have shown that the emission measure in \halpha\ is strongly correlated to the absolute intensity in \halpha, $I_{\mathrm{H}}$ (see Fig.~3 of their paper). This relation is expected to hold even for temperatures of $20,000~$K and heights of $30,000$~km, due to the dominance of photoionisation in the Lyman continuum and scattering of the incident radiation from the solar disk (the expected ionisation degree is still relatively low, on the order of 0.9999). We approximated this correlation with 2 linear functions. For $\log I_{\mathrm{H}}<4.979$ we take $EM_{\mathrm{H}}=\frac{\log I_{\mathrm{H}}-1.955}{1.008}+26$, and for $\log I_{\mathrm{H}}>4.979$ we take $EM_{\mathrm{H}}=\frac{\log I_{\mathrm{H}}-4.979}{0.536}+29$. 
	
	For a given LOS across a blob we assumed that the emergent \halpha\ intensity is produced by the combination of a background emission around the blob and the blob emission.
	For simplification we assumed that rain blobs are axisymmetric, so that their width along the LOS is the same as the one observed in the POS. In practice, we detected rain blobs by fitting multiple Gaussian profiles to the average intensity along the slices (see Fig.~\ref{fig:strands_sst}). We then detected the different rain strands that contains these blobs by locating the local intensity maxima along each parallel cut to the path. In that way we can differentiate the rain blob pixels from the background pixels (no blob fitted and under an intensity threshold). For a each blob pixel we then have:
	
	\begin{eqnarray}
	    EM_{\mathrm{H},bg+c}=\langle EM_{\mathrm{H},bg}\rangle+EM_{\mathrm{H},c},\\
		EM_{\mathrm{H},bg+c}=\langle EM_{\mathrm{H},bg}\rangle+n_{e,c}^2w,
	\end{eqnarray}

	where $n_{e,c}$ and $w$ are, respectively, the electron number densities and the width of the blobs. $\langle EM_{\mathrm{H},bg}\rangle$ denotes the average over all the background rain pixels for a given parallel cut. We then obtain:
	\begin{equation}\label{eq:density}
    	n_{e,c}=\sqrt{\frac{1}{w}\left(EM_{\mathrm{H},bg+c}-\langle EM_{\mathrm{H},bg}\rangle\right)},
	\end{equation}
	 If for some pixels $EM_{\mathrm{H},bg+c}$ happen to be smaller than $\langle EM_{\mathrm{H},bg}\rangle$, this pixel is discarded from the final calculation (in practice, it represents very few cases).
	 This density expression leads to an estimation for each rain pixel, in other words, on the LOS, rather than a density estimation for each blob.
	 This density estimation assumes that the rain blobs are optically thin in \halpha, an assumption that usually holds given the very small dimensions of the blobs and their not too high densities \citep[as suggested from numerical models,][]{fang_multidimensional_2013}\footnote{We checked this by taking a much thicker than usual rain blob of several Mm forming in an expanding 1D flux tube from the MHD model of \citet{antolin_coronal_2010}, embedding the flux tube in an atmosphere in hydrostatic and radiative equilibrium from \citet{Carlsson_2002ApJ...572..626C}, and performing the non-LTE calculation with the MULTI code to retrieve the \halpha\ line profile along various LOSs crossing the blob during its lifetime. The maximum optical thickness for this extreme case was found to be 0.5, thereby confirming that rain blobs are usually optically thin in \halpha.}.

\begin{figure*}
	\resizebox{\hsize}{!}
	{\includegraphics[trim={0cm 0cm 0cm 0cm},clip]{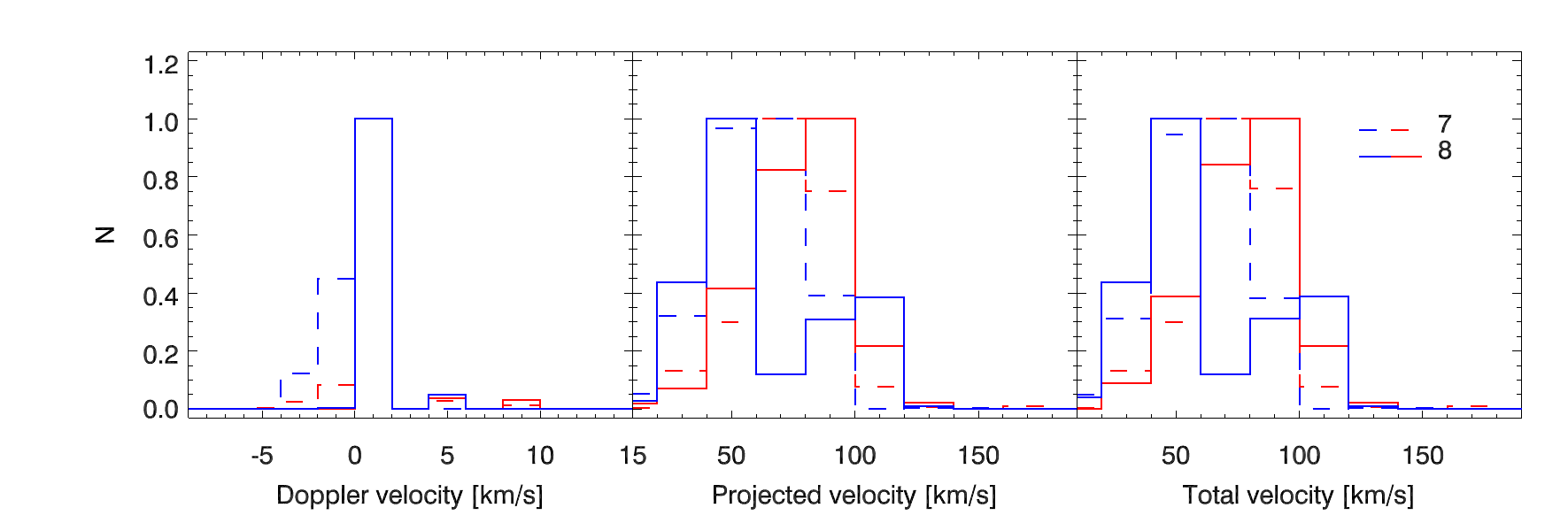}}
	\caption{Same as Fig.~\ref{fig:stats_vel} for paths 7 and 8 that are located in the core of the active region and are both covered by CRISP and CHROMIS.}
	\label{fig:stats_vel_other}
\end{figure*}	

\begin{figure*}
	\resizebox{\hsize}{!}
	{\includegraphics[trim={0cm 0cm 0cm 0.5cm},clip]{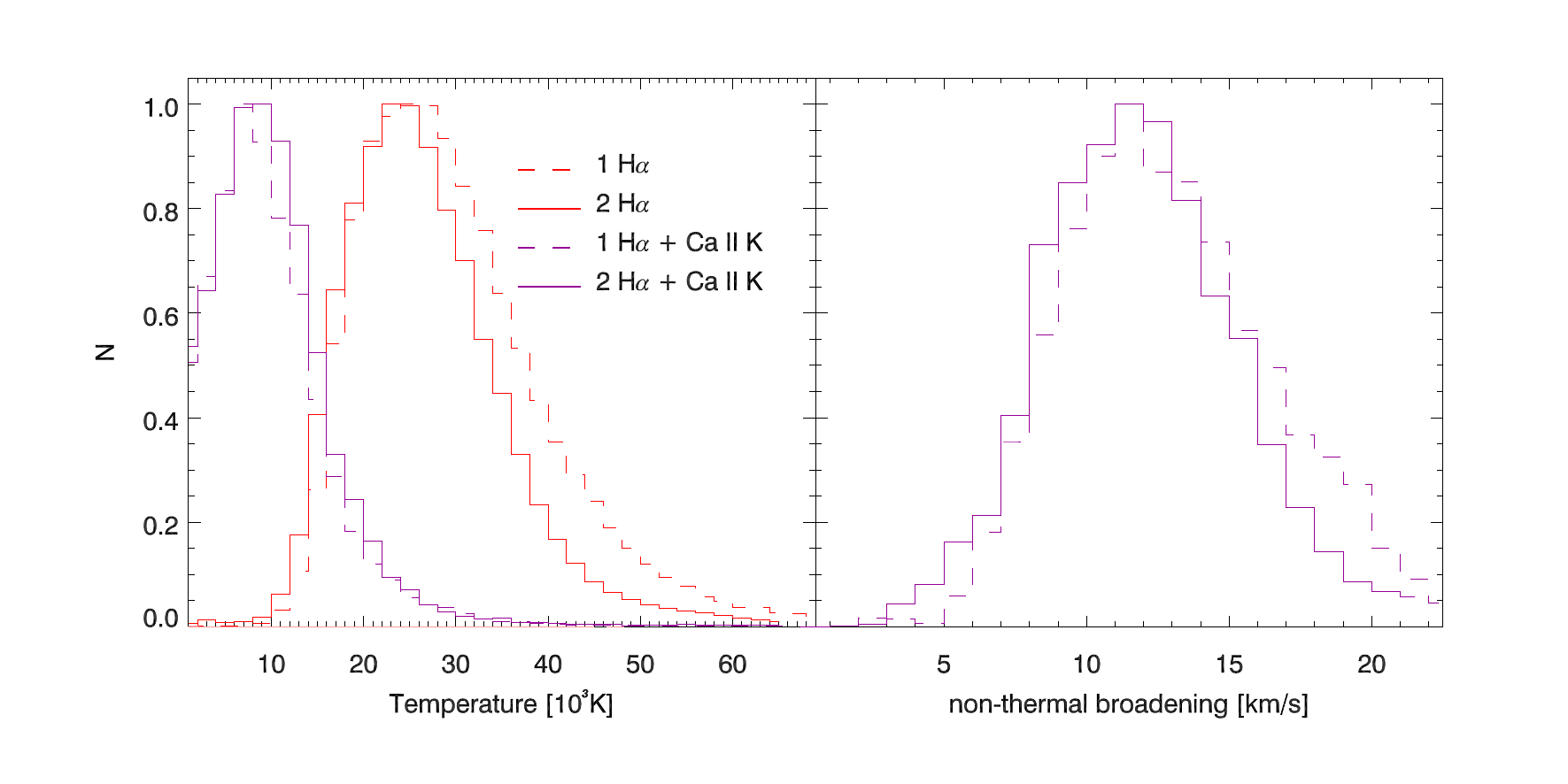}}
	\caption{Left: normalised histograms of the temperature estimated with CRISP only (\halpha, Equation~\ref{eq:Halpha_temp}) and the combination of CRISP and CHROMIS measurements (\halpha\ and  \ion{Ca}{II}~K, Equation~\ref{eq:T}) for paths 1 and 2. Right: normalised histograms of the estimation of the non-thermal broadening from Equation~\ref{eq:nth} for the same paths.  We only show the measurements that have an error below 15\% of the measured value.} 

	\label{fig:stats_temp}
\end{figure*}	

\begin{figure*}
	\resizebox{\hsize}{!}
	{\includegraphics[trim={0cm 0cm 0cm 0.5cm},clip]{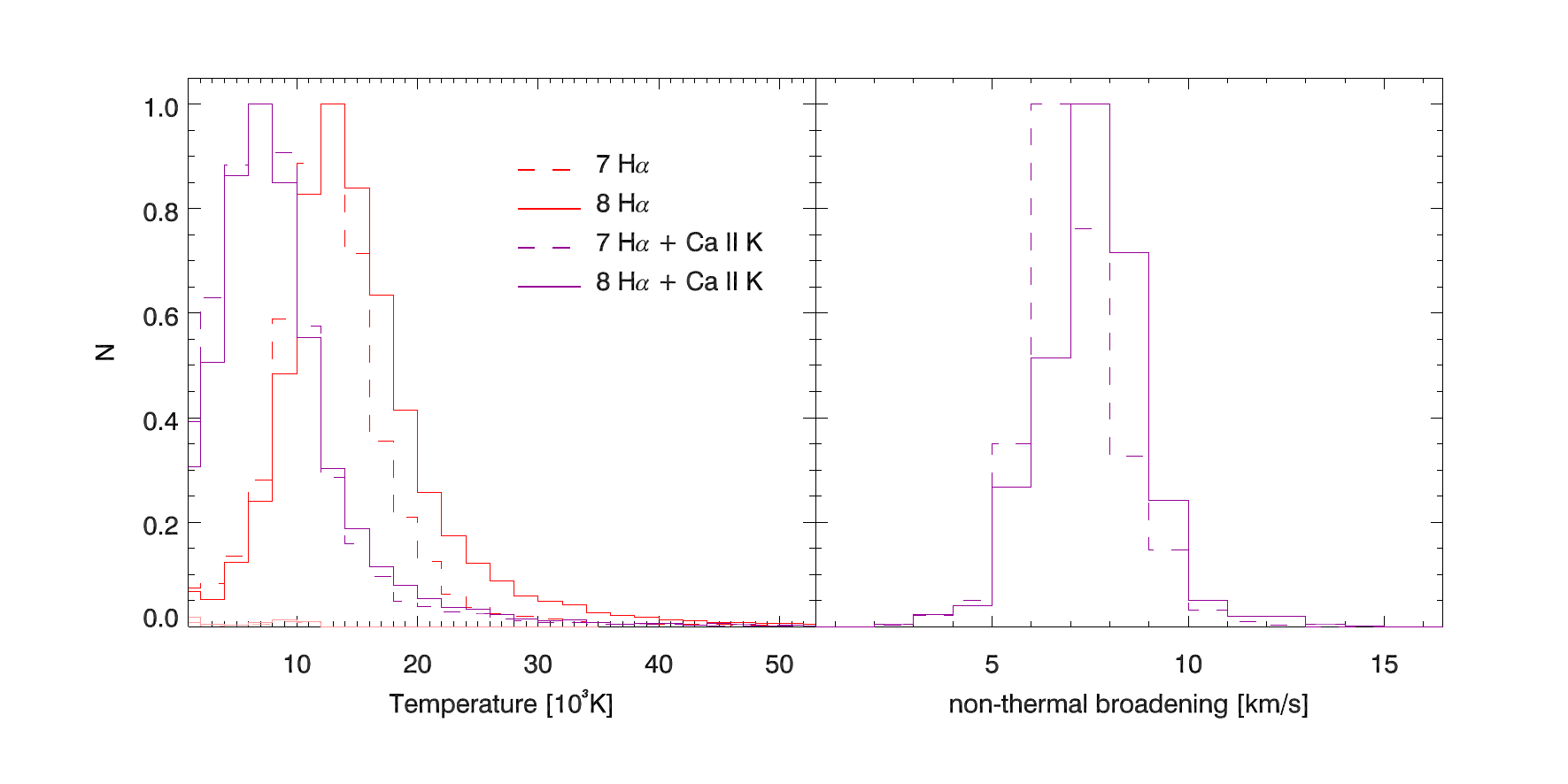}}
	\caption{Same as Fig.~\ref{fig:stats_temp} for paths 7 and 8.} 
	\label{fig:stats_temp_other}
\end{figure*}

\begin{figure}
	\centering
		\resizebox{\hsize}{!}
	{$\begin{array}{c}
		\includegraphics[trim={0cm 0cm 0cm 0cm},clip]{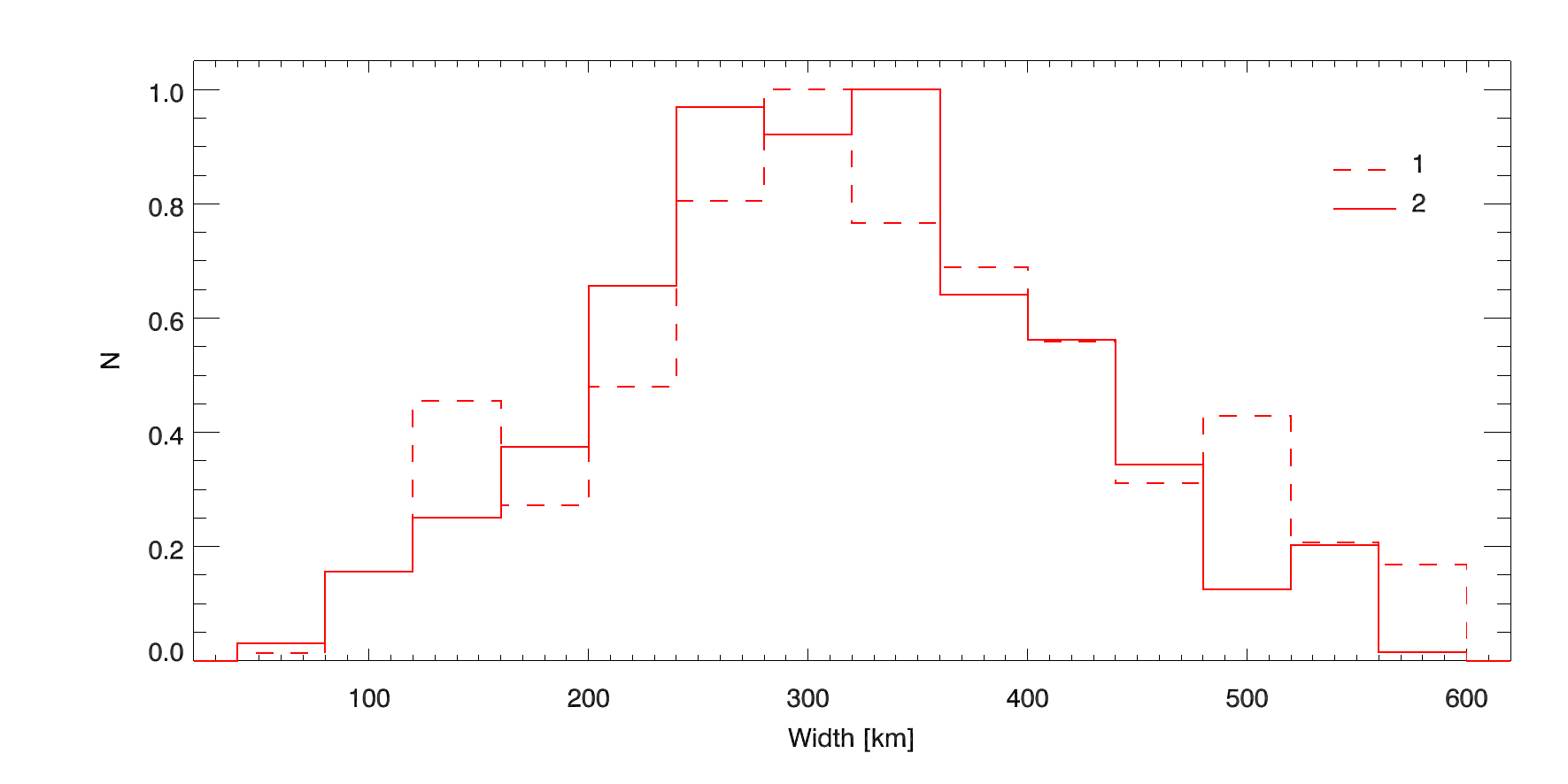} \\
		\includegraphics[trim={0cm 0cm 0cm 0cm},clip]{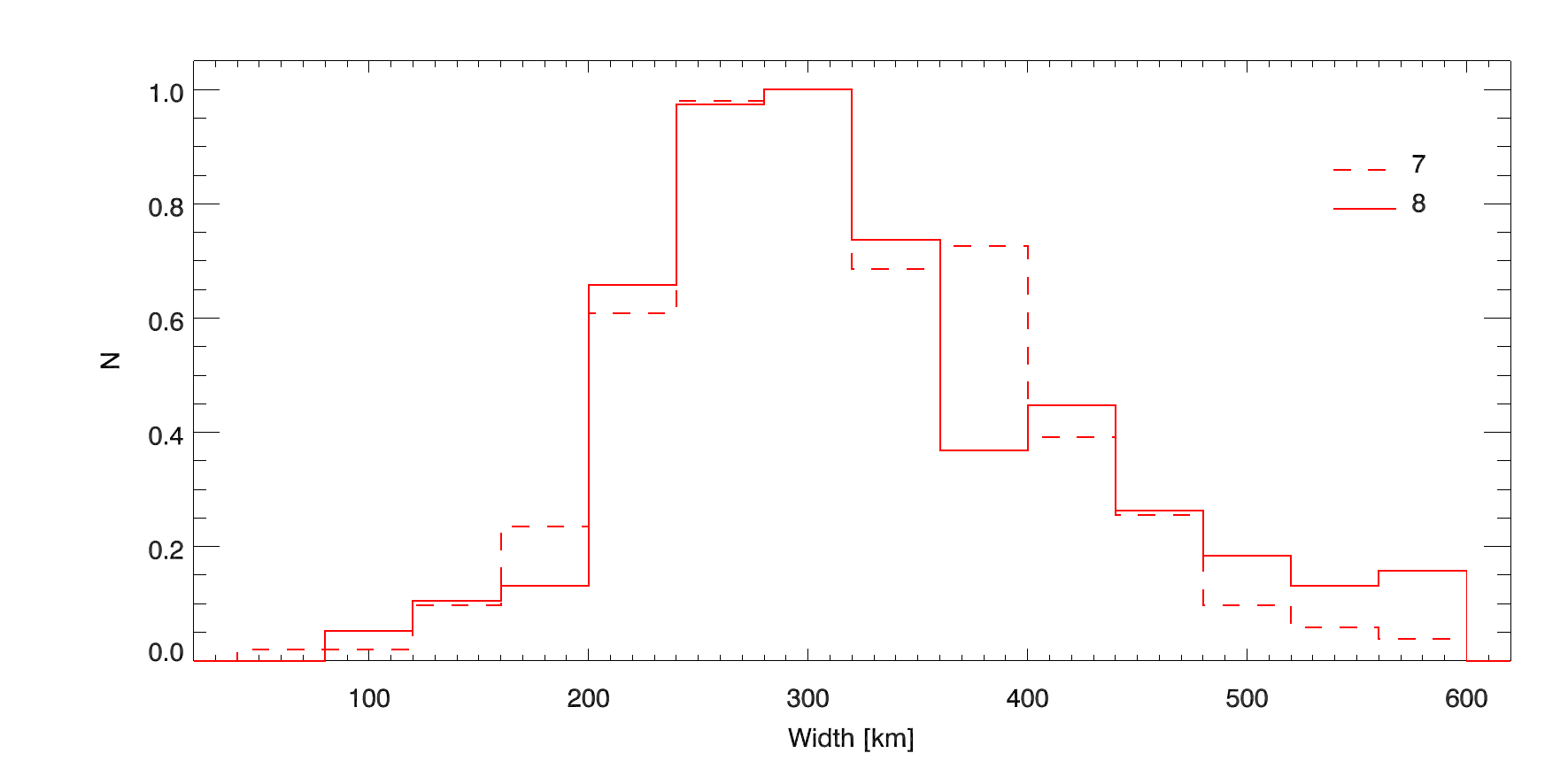}
	\end{array}$
	}
	\caption{Normalised histograms of the widths of the rain blobs determined for the density calculations from \halpha. These widths are determined as explained in Sect.~\ref{sec:method_rain} and shown in Fig.~\ref{fig:strands_sst}.  We only show the measurements that have an error below 15\% of the measured value. Top: Paths 1 and 2. The total number of rain blobs detected are, respectively, 533 and 450. Bottom: Paths 7 and 8. The total number of rain blobs detected are, respectively, 283 and 213. 
	}
	\label{fig:stats_width}
\end{figure}	

\begin{figure*}
	\resizebox{\hsize}{!}
	{\includegraphics[trim={0cm 0cm 0cm 0cm},clip]{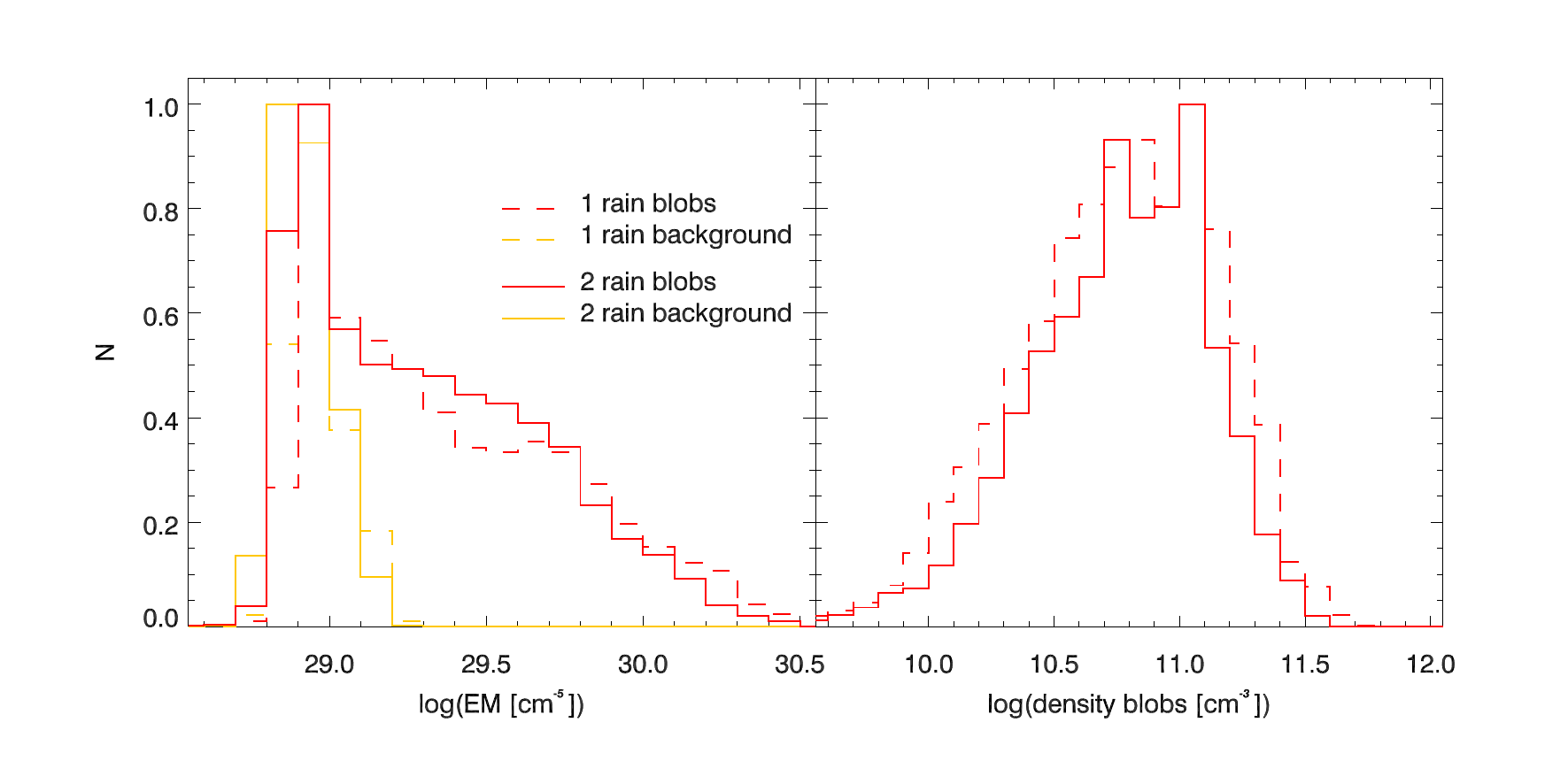}}
	\caption{Normalised histograms of the EM (from the rain blobs: $EM_{\mathrm{H},c}$ and the rain background: $\langle EM_{\mathrm{H},bg}\rangle$) and density measured from the \halpha\ intensities (see Method in Sect.~\ref{sec:method_rain}).} 
	\label{fig:stats_density}
\end{figure*}	

\begin{figure*}
	\resizebox{\hsize}{!}
	{\includegraphics[trim={0cm 0cm 0cm 0cm},clip]{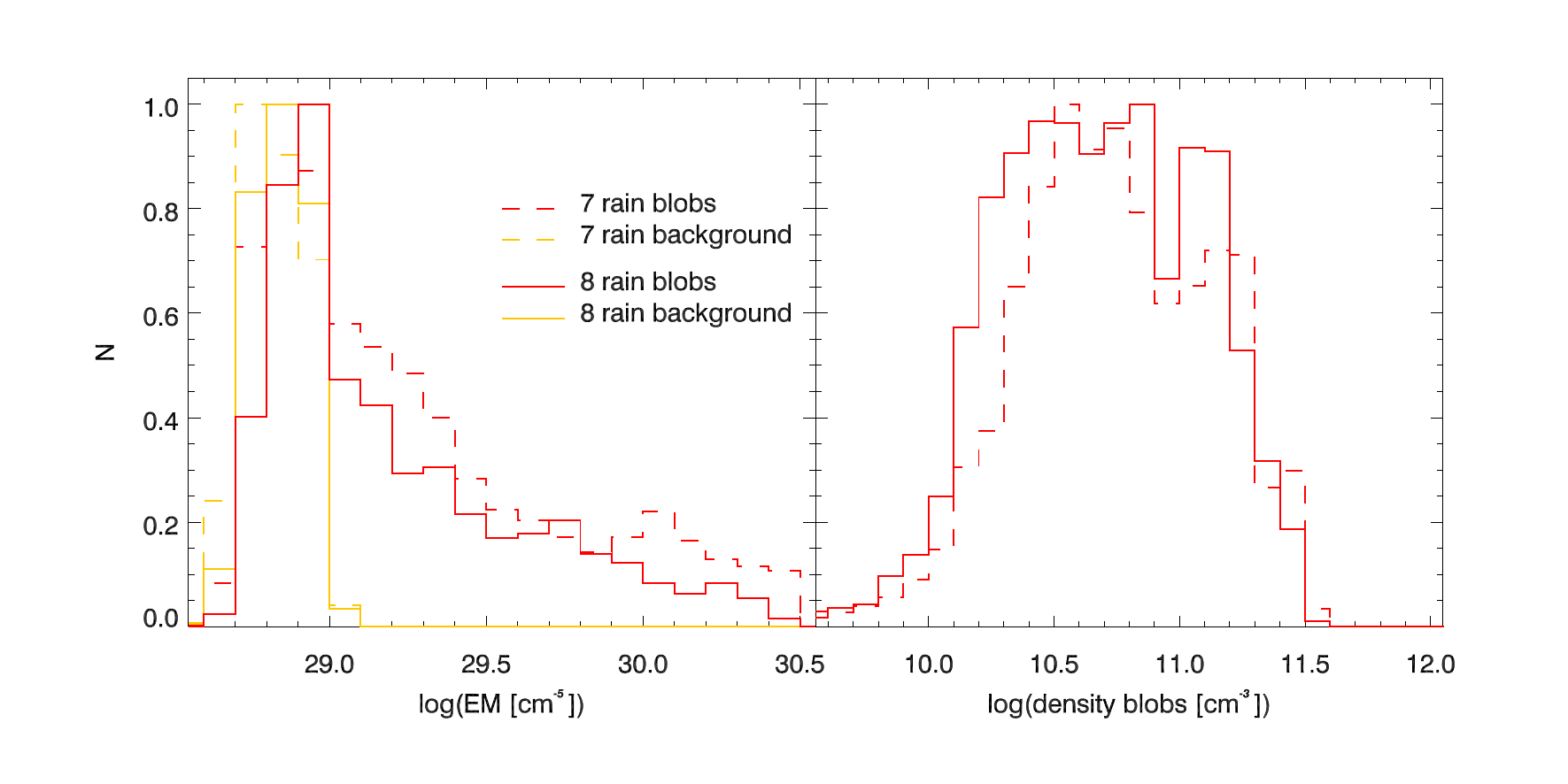}}
	\caption{Same as Fig.~\ref{fig:stats_density} for paths 7 and 8.}
	\label{fig:stats_density_other}
\end{figure*}	

\subsubsection{Results}\label{sec:results_rain}

For this analysis we chose two sets of paths. The first set covers two different range of altitudes in the pulsating loop bundle, namely, paths 1 to 6 displayed in the left panel of Fig.~\ref{fig:sst_path}. Among this first set of paths, if all of them are covered by the CHROMIS FOV, only 1 and 2 are covered by the CRISP FOV. These two paths are then the only ones used for the temperature and non-thermal line broadening measurements and the density measurements. Since paths 3 and 4 are only serving as a junction between paths 1 and 5, and 2 and 6, respectively, we combine them with paths 5 and 6 for the velocity analysis.
The second set covers two different ranges of altitudes in a bundle located in the core of the active region, that is, paths 7 to 8 displayed in the right panel of Fig.~\ref{fig:sst_path}. This second set is used as a test case to be compared with our estimations for paths 1 to 6.

\paragraph{Velocity.}

The velocity statistics of the rain blobs analysed in path 1 to 6 and paths 7 and 8 are displayed in  Fig.~\ref{fig:stats_vel} and~\ref{fig:stats_vel_other}, respectively.  These distributions are constructed from velocity measurements for which the error is less than 15$\%$ of the measured value.
For paths 1 to 6, Doppler velocities range from 20 to 60~\kms. For path 1 and 2 the average is, respectively, 40 and 37~\kms\ 
from \halpha\ and 36 and 35~\kms\ from \ion{Ca}{II}~K. 
Similar Doppler velocities are derived from paths 3 to 6 (34 and 27~\kms\ in average). 
As for the temperature measurements no major differences can be noted between the two ranges of altitudes covered by the different paths. The projected velocity, and thus the total velocity, distributions are quite broad, with velocities ranging from about 30 to 150~\kms. The average projected velocities are, respectively, 77 and 85~\kms\  
for path 1 and 2 with CRISP and 73 and 88~\kms\
for path 1 and 2 with CHROMIS and  61 and 70~\kms\ 
for path 3\&5 and 4\&6. The total velocities are, respectively, 88 and 94~\kms\ 
for path 1 and 2 with CRISP and 84 and 99~\kms\ 
for path 1 and 2 with CHROMIS and 82 and 88~\kms\ 
for path 3\&5 and 4\&6.

Paths 7 and 8, that are located in the core of the active region (non-pulsating loops), show rain blobs with smaller velocities. The Doppler velocities range from -5 to 10~\kms\, with an average of about 0~\kms\  
and for both \halpha\ and \ion{Ca}{II}~K. 
Projected velocities range from 30 to 130~\kms\, with an average of, respectively, 74 and 59~\kms\ 
 with CRISP and 76 and 62~\kms\ 
with CHROMIS. The total velocities are, respectively, 74 and 76~\kms\ 
with CRISP and 60 and 64~\kms\ 
with CHROMIS.

In summary, the velocity distributions for \halpha\ and \ion{Ca}{II}~K are generally similar for the same path.
Moreover, the two sets of paths, namely, in the pulsating loop bundle and in the core of the active region, are showing similar total velocities.

\paragraph{Temperature and non-thermal line broadening.}

The temperatures and non-thermal broadening statistics for paths 1 and 2, and paths 7 and 8 are presented in, respectively, Fig.~\ref{fig:stats_temp} and Fig.~\ref{fig:stats_temp_other}.  As for the velocity distributions, the temperature distributions are constructed from measurements for which the error is less than 15$\%$ of the measured value. We note that for path 1 and 2, respectively, 76\% and 82\% of the rain pixels satisfy the condition $\frac{\sigma_{\mathrm{H}}}{\lambda_{0,\mathrm{H}}}>\frac{\sigma_{\mathrm{Ca}}}{\lambda_{0,\mathrm{Ca}}}$, that is true if we indeed observe the same element of plasma. It increases to 92\% and 95\% for path 7 and 8. It is thus possible to analyse most of the detected rain pixels. Paths 3 to 6 are not covered by the CRISP FOV and no such estimation are possible with only \ion{Ca}{II}~K as developed in the Sect.~\ref{sec:method_rain}.
In these figures we show both the temperature estimation from \halpha\ only and for the combination of both \halpha\ and \ion{Ca}{II}~K.

From \halpha\ only, the average temperature for paths 1 and 2 is, respectively, $\sim 30.5 \times 10^3$~K and $27.5  \times 10^3$~K, 
while for paths 7 and 8 the average temperature is, respectively, $\sim 13.5 \times 10^3$~K and $\sim 16 \times 10^3$~K. 
For the first set of paths we also notice that the distribution has an extended tail up to $60 \times 10^3$ K. Such a tail is also present in the temperatures computed for the second set of paths (extending toward $40 \times 10^3$ K).

From the combination of both \halpha\ and \ion{Ca}{II}~K, these temperature distributions seem to conserve their shape (i.e. log-normal). However, the average temperature drops by about $20 \times 10^3$ K for paths 1 and 2 (for each path compared to the measurement from \halpha\ only), while the decrease of the average temperature is of about $5 \times 10^3$ K for paths 7 and 8. In fact, the non-thermal broadening is in average bigger for paths 1 and 2, respectively, 13~\kms\ and 12~\kms, than for paths 7 and 8, respectively, 7~\kms\ and 8~\kms.  
The average temperature with this method for paths 1 and 2 is, respectively, $\sim10 \times 10^3$~K and $10.5  \times 10^3$~K, for paths 7 and 8 the average temperature drops to, respectively, $\sim 8 \times 10^3$~K and $\sim9.5 \times 10^3$~K.

The following points summarise the characteristics we observed. Firstly, the temperature and non-thermal broadening distributions are similar between the two ranges of altitudes  covered, that is, between, respectively, paths 1 and 2, and paths 7 and 8.
Secondly, the temperature distribution is shifted toward smaller temperatures when we combine \halpha\ and \ion{Ca}{II}~K. Since the non-thermal broadening is no longer neglected and even estimated by this method, we no longer take into account this extra broadening as a thermal component.
However, the temperatures found for both sets of loops are similar when \halpha\ and \ion{Ca}{II}~K are combined. 
Finally, very low temperatures seem to be reached ($\sim 2000$~K). They seem to originate from spatially coherent areas even though further investigations are needed to understand their origin (and whether they are real or an artefact of the method applied to these particular data).

As we have seen in the velocity analysis, the Doppler velocity distributions for paths 1 and 2 have a larger standard deviation than the one for paths 7 and 8. A large Doppler distribution can result in the imprint of multiple blobs in the spectra, affecting the width of the Gaussian fit and thus resulting in higher temperature estimations. Even if we tried to discard as much as possible such profiles from the statistics, it is not possible to always distinguish them. Thus, it can explain why the temperature estimation from \halpha\ only is higher for paths 1 and 2 than for paths 7 and 8 and why the temperature difference is much bigger for paths 1 and 2 than for paths 7 and 8 when \halpha\ and \ion{Ca}{II}~K are combined (similarly, the non-thermal broadening is more important for the first set of paths). 
The fact that we combined the widths measurements of two lines to compute the temperatures (see Equation~\ref{eq:T}) might help to reduce the imprint of multiple blobs in the spectra, although such effect is difficult to disentangle from the non-thermal broadening that is due to micro-turbulence.

The difference in non-thermal broadening for these two loop bundles could be due to their different geometry (the first one extended along the LOS and the other one mostly in the POS), as developed above. Further investigations would be required to determine whether other effects are at play, such as the position in the core of the active region of the loops covered by the second set of paths.

\paragraph{Blobs widths.}

The width distribution of the rain blobs detected in \halpha, extracted for the density measurements (see Sect.~\ref{sec:method_rain}), are displayed in Fig.~\ref{fig:stats_width} for paths 1 and 2, and 7 and 8. 
The width distributions are quite similar for both bundles and the range of altitudes covered. The widths extracted range from 100 to 600 km, with an average of  about 330 km for all the paths (from measurements which error is less than 15$\%$ of the measured value.

\paragraph{Density.}

Figures~\ref{fig:stats_density} and~\ref{fig:stats_density_other} show the distribution of the emission measure of the background rain pixels $\langle EM_{\mathrm{H},bg}\rangle$ and the emission measure for the rain pixels $EM_{\mathrm{H},bg+c}$, as well as the subsequent density estimation (Equation~\ref{eq:density}) for, respectively, paths 1 and 2, and 7 and 8.
For both sets of paths, the distribution of the emission measures are similar: a peak around $10^{29}$~cm$^{-5}$, that is located in the same range of EM as the background rain pixels, and a large tail that goes up to $10^{30.5}$~cm$^{-5}$.

The density distributions range from $10^{9}$~cm$^{-3}$ to $10^{12}$~cm$^{-3}$. The distributions for path 1 and 2 are similar, with an average of $8.5 \times 10^{10}$~cm$^{-3}$ and $7.8 \times 10^{10}$~cm$^{-3}$, respectively. 
However, while the distributions for paths 7 and 8 are also similar, they seem to show two distinct peaks (one around $10^{10.7}$~cm$^{-3}$ and the other around $10^{11.5}$~cm$^{-3}$), that were not present for paths 1 and 2. The average density is, respectively, $8.0 \times 10^{10}$~cm$^{-3}$ and $6.7 \times 10^{10}$~cm$^{-3}$. 
For these density distributions, we did not derive an estimation of the errors. The uncertainties coming from the calibration of the absolute intensity and the model used, which may be the main factors, are difficult to estimate. We thus prefer not to underestimate or overestimate them.

\subsubsection{Coronal rain properties and contribution to the chromosphere-corona mass cycle}

The coronal rain blobs we analysed seem to have similar properties, whether they originate from a bundle for which we detect periodic TNE cycles (paths 1 to 6) or not (paths 7 and 8).
Their kinetic and thermodynamic properties are also consistent with other coronal rain studies.
The velocities, temperatures and blob widths found are comparable with those of other observational rain studies \citep[e.g.][]{de_groof_detailed_2005, Ahn_2014SoPh..289.4117A, Antolin_Rouppe_2012ApJ...745..152A, Antolin_2015ApJ...806...81A, Kohutova_2016ApJ...827...39K, Schad_2017SoPh..292..132S}, indicating that the rain observed here, together with long-period intensity pulsations, is consistent with the classical so-called quiescent coronal rain that has been observed for decades.

The mass drain rate found here from the density measurements is also comparable with other studies.
For path 1 and 2 we use, respectively, an average mass density of 1.4 and $1.3 \times 10^{-13}$~g cm$^{-3}$ and for path 7 and 8, respectively, 1.3 and $1.1 \times 10^{-13}$~g cm$^{-3}$.
Taking in account the number of blobs and their average width described earlier, an average length of 957~km (the width covered by our perpendicular path, since the blobs seem to always cover most of this length). We obtain a mass rate of 3.5 and $2.8 \times 10^{9}$~g s$^{-1}$ for path 1 and 2, 1.4 and $1.3 \times 10^{9}$~g s$^{-1}$ for path 7 and 8.
This is similar to what was found for other coronal rain studies \citep[e.g.][]{Antolin_2015ApJ...806...81A} as well as from prominence drainage \citep[e.g.][]{liu_first_2012}.

\section{Multi-scale TNE cycles}    

In this paper, we analysed a TNE event combining SDO and SST observations that allow us to probe one cooling phase in its entirety. For the first time, we were able to cover the extreme scales involved in TNE events, that have long been predicted by numerical simulations. 

On one hand, the active region, observed off-limb for three days, is showing 6-hour periodic pulsations in most of the AIA channels, with very strong signal\footnote{For the sake of conciseness and because another long-period intensity pulsation event off-limb has already been studied in detail in \citet{auchere_coronal_2018}, we chose not to explore the pulse-train nature of these signals in this paper. However, because of the shape of the Fourier spectra, we would expect to find the same signatures as detailed in \citet{auchere_thermal_2016}.}. These pulsations are located mostly at the apex of a large loop bundle. Within the same bundle, coronal rain is periodically formed near the loops' 
apex and evacuated towards both legs. It seems that the rain appears first in the internal loops of the bundle and then at the very top of the bundle. Similar behaviour was shown by 2D numerical simulations \citep[e.g.][]{fang_multidimensional_2013, Fang_2015ApJ...807..142F}, but also by 1D simulations since, for similar heating conditions, the longer the loop the longer the period, explaining the observed time delay and the visual gradual expansion of the coronal loop bundle (C. Downs, private communication). The DEM analysis revealed that the thermal structure of the bundle undergoes a periodic modification of the quantity of cool plasma along the LOS. Moreover, the density evolves periodically and is delayed compared to the temperature evolution. The 1.1 hour time lag found between the temperature and the density corresponds to 18\% of the period and is therefore compatible with the range of 20-30\% for TNE cases, with coronal rain or incomplete condensations, found in \citet{froment_occurrence_2018} with numerical simulations.
This loop bundle cools through the temperature range of 2.5~MK-0.08~MK, as pointed out by a time-lag analysis.
In conclusion, we present clear evidence that this loop bundle undergo TNE cycles and that the thermodynamics of this region can be thus explained by a quasi-steady and stratified heating.

On the other hand, we studied in detail the full catastrophic cooling of one of these cycles. Using the CRISP and CHROMIS instruments at SST, we observed coronal blobs, 300~km wide in average, with temperatures of about $10\,000$~K and densities of about $8 \times 10^{10} \; \mathrm{cm}^{-3}$. These coronal rain blobs are falling in clumps at an average velocity of about $90 \; \mathrm{km} \; \mathrm{s}^{-1}$. These characteristics are very typical of coronal rain observations and simulations. We estimated the mass drain rate to be about $3 \times 10^{9}$~g s$^{-1}$, reaffirming the role of coronal rain in the corona-chromosphere mass cycle.

These observations confirm the global behaviour of TNE, connecting the many different temporal and spatial scales of the evolution of temperature and density conditions, and its role in structuring the dynamics of the plasma from the corona to the chromosphere. On top of the loop bundle undergoing periodic TNE cycles as studied here, other loops located in the core of the active region also show regular coronal rain events but no periodicity.
Quasi-constant footpoint heating can also explain the rain formation in the core of the active region but it might be that the variability of the heating parameter, and thus variable cycles may lead to no intrinsic periodicity. If the TNE conditions are broken after one cycle or two, no detection would be possible. Furthermore, the LOS superimposition of different loops is likely to make cycles undetectable.

Some other loops did not seem to show any signs of TNE (no pulsations and no rain). However, stratified and high frequency heating can be present beyond structures that are experiencing TNE, as shown in \citet{froment_occurrence_2018}. In some cases, close to TNE-prone conditions, the loops global behaviour can be dominated by siphon flows without thermal cycles \citep[ also see][]{klimchuk_role_2019}. 
Future studies ought to determine if the spatiotemporal structure of the heating in active regions, such as the one studied in the present paper, can be understood by a distribution of stratifications and heating frequencies.

\begin{acknowledgements}
	
	The authors thank S. Jafarzadeh for his help during the co-alignment between CRISP and AIA and M. Szydlarski for his help for the parallelization of the DEM calculation. We also thank J. -C. Vial for fruitful discussions on the density determination method.  We thank T. M. Chamandy who was co-observer together with V. M. J. Henriques and C. Froment at the Swedish 1-m Solar Telescope (SST) during this campaign and, H. Pazira and P. S\"{u}tterlin for technical support at the SST. The Swedish 1-m Solar Telescope is operated on the island of La Palma by the Institute for Solar Physics of Stockholm University in the Spanish Observatorio del Roque de los Muchachos of the Instituto de Astrof\'{i}sica de Canarias. The Institute for Solar Physics is supported by a grant for research infrastructures of national importance from the Swedish Research Council (registration number 2017-00625). The SDO/AIA are available courtesy of NASA/SDO and the AIA science teams. This work used data provided by the MEDOC data and operations center (CNES/CNRS/Univ. Paris-Sud), http://medoc.ias.u-psud.fr/.  This research was supported by the Research Council of Norway, project no. 250810, and through its Centres of Excellence scheme, project no. 262622. P. Antolin acknowledges funding from his STFC Ernest Rutherford Fellowship (No. ST/R004285/1).
	C. Froment and P. Antolin also acknowledge support from the International Space Science Institute (ISSI), Bern, Switzerland as well as fruitful discussions with members of the International Team 401 \lq Observed Multi-Scale Variability of Coronal Loops as a Probe of Coronal Heating\rq. 
	
\end{acknowledgements}

\bibliographystyle{aa}
\bibliography{Refs_AtoE,Refs_FtoJ,Refs_KtoO,Refs_PtoT,Refs_UtoZ,patbib}

\begin{thebibliography}{90}
\expandafter\ifx\csname natexlab\endcsname\relax\def\natexlab#1{#1}\fi

\bibitem[{{Ahn} {et~al.}(2014){Ahn}, {Chae}, {Cho}, {Song}, {Yang}, {Goode},
  {Cao}, {Park}, {Nah}, {Jang}, \& {Park}}]{Ahn_2014SoPh..289.4117A}
{Ahn}, K., {Chae}, J., {Cho}, K.-S., {et~al.} 2014, \solphys, 289, 4117

\bibitem[{{Allen}(1973)}]{Allen1973asqu.book.....A}
{Allen}, C.~W. 1973, {Astrophysical quantities}

\bibitem[{Antiochos \& Klimchuk(1991)}]{antiochos_model_1991}
Antiochos, S.~K. \& Klimchuk, J.~A. 1991, The Astrophysical Journal, 378, 372

\bibitem[{{Antiochos} {et~al.}(2000){Antiochos}, {MacNeice}, \&
  {Spicer}}]{Antiochos_2000ApJ...536..494A}
{Antiochos}, S.~K., {MacNeice}, P.~J., \& {Spicer}, D.~S. 2000, \apj, 536, 494

\bibitem[{{Antiochos} {et~al.}(1999){Antiochos}, {MacNeice}, {Spicer}, \&
  {Klimchuk}}]{Antiochos_1999ApJ...512..985A}
{Antiochos}, S.~K., {MacNeice}, P.~J., {Spicer}, D.~S., \& {Klimchuk}, J.~A.
  1999, \apj, 512, 985

\bibitem[{Antolin(2019)}]{antolin_thermal_2019}
Antolin, P. 2019, Plasma Physics and Controlled Fusion

\bibitem[{{Antolin} \& {Rouppe van der
  Voort}(2012)}]{Antolin_Rouppe_2012ApJ...745..152A}
{Antolin}, P. \& {Rouppe van der Voort}, L. 2012, \apj, 745, 152

\bibitem[{Antolin {et~al.}(2010)Antolin, Shibata, \&
  Vissers}]{antolin_coronal_2010}
Antolin, P., Shibata, K., \& Vissers, G. 2010, The Astrophysical Journal, 716,
  154

\bibitem[{{Antolin} {et~al.}(2015){Antolin}, {Vissers}, {Pereira}, {Rouppe van
  der Voort}, \& {Scullion}}]{Antolin_2015ApJ...806...81A}
{Antolin}, P., {Vissers}, G., {Pereira}, T.~M.~D., {Rouppe van der Voort}, L.,
  \& {Scullion}, E. 2015, \apj, 806, 81

\bibitem[{{Antolin} {et~al.}(2012){Antolin}, {Vissers}, \& {Rouppe van der
  Voort}}]{Antolin_etal_2012SoPh..280..457A}
{Antolin}, P., {Vissers}, G., \& {Rouppe van der Voort}, L. 2012, \solphys,
  280, 457

\bibitem[{Auch{\`e}re {et~al.}(2014)Auch{\`e}re, Bocchialini, Solomon, \&
  Tison}]{auchere_long-period_2014}
Auch{\`e}re, F., Bocchialini, K., Solomon, J., \& Tison, E. 2014, \aap, 563, A8

\bibitem[{Auch{\`e}re {et~al.}(2016{\natexlab{a}})Auch{\`e}re, Froment,
  Bocchialini, Buchlin, \& Solomon}]{auchere_fourier_2016}
Auch{\`e}re, F., Froment, C., Bocchialini, K., Buchlin, E., \& Solomon, J.
  2016{\natexlab{a}}, The Astrophysical Journal, 825, 110

\bibitem[{Auch{\`e}re {et~al.}(2016{\natexlab{b}})Auch{\`e}re, Froment,
  Bocchialini, Buchlin, \& Solomon}]{auchere_thermal_2016}
Auch{\`e}re, F., Froment, C., Bocchialini, K., Buchlin, E., \& Solomon, J.
  2016{\natexlab{b}}, The Astrophysical Journal, 827, 152

\bibitem[{Auch{\`e}re {et~al.}(2018)Auch{\`e}re, Froment, Soubri{\'e}, Antolin,
  Oliver, \& Pelouze}]{auchere_coronal_2018}
Auch{\`e}re, F., Froment, C., Soubri{\'e}, E., {et~al.} 2018, The Astrophysical
  Journal, 853, 176

\bibitem[{Barnes {et~al.}(2019)Barnes, Bradshaw, \&
  Viall}]{barnes_understanding_2019}
Barnes, W.~T., Bradshaw, S.~J., \& Viall, N.~M. 2019, The Astrophysical
  Journal, 880, 56

\bibitem[{Boerner {et~al.}(2012)Boerner, Edwards, Lemen, Rausch, Schrijver,
  Shine, Shing, Stern, Tarbell, Title, Wolfson, Soufli, Spiller, Gullikson,
  McKenzie, Windt, Golub, Podgorski, Testa, \& Weber}]{boerner_initial_2012}
Boerner, P., Edwards, C., Lemen, J., {et~al.} 2012, \solphys, 275, 41

\bibitem[{Bradshaw \& Cargill(2010)}]{bradshaw_cooling_2010}
Bradshaw, S.~J. \& Cargill, P.~J. 2010, The Astrophysical Journal, 717, 163

\bibitem[{Bradshaw \& Viall(2016)}]{bradshaw_patterns_2016}
Bradshaw, S.~J. \& Viall, N.~M. 2016, The Astrophysical Journal, 821

\bibitem[{Brault \& Neckel(1987)}]{brault_atlas}
Brault, J.~W. \& Neckel, H. 1987, Spectral Atlas of Solar Absolute
  Disk-averaged and Disk-Center Intensity from 3290 to 12510 $\AA$,
  \url{ftp://ftp.hs.uni-hamburg.de/pub/outgoing/FTS-Atlas}

\bibitem[{{Carlsson} \& {Stein}(2002)}]{Carlsson_2002ApJ...572..626C}
{Carlsson}, M. \& {Stein}, R.~F. 2002, \apj, 572, 626

\bibitem[{Cauzzi {et~al.}(2009)Cauzzi, Reardon, Rutten, Tritschler, \&
  Uitenbroek}]{cauzzi_solar_2009}
Cauzzi, G., Reardon, K., Rutten, R.~J., Tritschler, A., \& Uitenbroek, H. 2009,
  Astronomy \& Astrophysics, 503, 577

\bibitem[{Cheung {et~al.}(2015)Cheung, Boerner, Schrijver, Testa, Chen, Peter,
  \& Malanushenko}]{cheung_thermal_2015}
Cheung, M. C.~M., Boerner, P., Schrijver, C.~J., {et~al.} 2015, The
  Astrophysical Journal, 807, 143

\bibitem[{De~Groof {et~al.}(2005)De~Groof, Bastiaensen, M{\"u}ller, Berghmans,
  \& Poedts}]{de_groof_detailed_2005}
De~Groof, A., Bastiaensen, C., M{\"u}ller, D. A.~N., Berghmans, D., \& Poedts,
  S. 2005, Astronomy and Astrophysics, 443, 319

\bibitem[{{De Groof} {et~al.}(2004){De Groof}, {Berghmans}, {van
  Driel-Gesztelyi}, \& {Poedts}}]{DeGroof_2004AA...415.1141D}
{De Groof}, A., {Berghmans}, D., {van Driel-Gesztelyi}, L., \& {Poedts}, S.
  2004, \aap, 415, 1141

\bibitem[{{de la Cruz Rodr{\'{\i}}guez} {et~al.}(2013){de la Cruz
  Rodr{\'{\i}}guez}, {Rouppe van der Voort}, {Socas-Navarro}, \& {van
  Noort}}]{Jaime2013A&A...556A.115D}
{de la Cruz Rodr{\'{\i}}guez}, J., {Rouppe van der Voort}, L., {Socas-Navarro},
  H., \& {van Noort}, M. 2013, \aap, 556, A115

\bibitem[{de~la Cruz~Rodríguez {et~al.}(2015)de~la Cruz~Rodríguez, Löfdahl,
  Sütterlin, Hillberg, \& Rouppe van~der
  Voort}]{de_la_cruz_rodriguez_crispred:_2015}
de~la Cruz~Rodríguez, J., Löfdahl, M.~G., Sütterlin, P., Hillberg, T., \&
  Rouppe van~der Voort, L. 2015, Astronomy and Astrophysics, 573, A40

\bibitem[{Dere {et~al.}(2019)Dere, Zanna, Young, Landi, \&
  Sutherland}]{dere_chiantiatomic_2019}
Dere, K.~P., Zanna, G.~D., Young, P.~R., Landi, E., \& Sutherland, R.~S. 2019,
  The Astrophysical Journal Supplement Series, 241, 22

\bibitem[{Fang {et~al.}(2013)Fang, Xia, \&
  Keppens}]{fang_multidimensional_2013}
Fang, X., Xia, C., \& Keppens, R. 2013, The Astrophysical Journal Letters, 771,
  L29

\bibitem[{{Fang} {et~al.}(2013){Fang}, {Xia}, \&
  {Keppens}}]{Fang_2013ApJ...771L..29F}
{Fang}, X., {Xia}, C., \& {Keppens}, R. 2013, \apjl, 771, L29

\bibitem[{{Fang} {et~al.}(2015){Fang}, {Xia}, {Keppens}, \& {Van
  Doorsselaere}}]{Fang_2015ApJ...807..142F}
{Fang}, X., {Xia}, C., {Keppens}, R., \& {Van Doorsselaere}, T. 2015, \apj,
  807, 142

\bibitem[{{Field}(1965)}]{Field_1965ApJ...142..531F}
{Field}, G.~B. 1965, \apj, 142, 531

\bibitem[{{Froment}(2016)}]{2016PhDT.......115F}
{Froment}, C. 2016, PhD thesis, Universit{\'e} Paris-Saclay, Universit{\'e}
  Paris-Sud, Institut d'Astrophysique Spatiale, Orsay, France

\bibitem[{Froment {et~al.}(2017)Froment, Auch{\`e}re, Aulanier, Miki{\'c},
  Bocchialini, Buchlin, \& Solomon}]{froment_long-period_2017}
Froment, C., Auch{\`e}re, F., Aulanier, G., {et~al.} 2017, The Astrophysical
  Journal, 835, 272

\bibitem[{Froment {et~al.}(2015)Froment, Auch{\`e}re, Bocchialini, Buchlin,
  Guennou, \& Solomon}]{froment_evidence_2015}
Froment, C., Auch{\`e}re, F., Bocchialini, K., {et~al.} 2015, The Astrophysical
  Journal, 807, 158

\bibitem[{Froment {et~al.}(2018)Froment, Auch{\`e}re, Miki{\'c}, Aulanier,
  Bocchialini, Buchlin, Solomon, \& Soubri{\'e}}]{froment_occurrence_2018}
Froment, C., Auch{\`e}re, F., Miki{\'c}, Z., {et~al.} 2018, The Astrophysical
  Journal, 855, 52

\bibitem[{{Gouttebroze} {et~al.}(1993){Gouttebroze}, {Heinzel}, \&
  {Vial}}]{Gouttebroze_etal_1993AAS...99..513G}
{Gouttebroze}, P., {Heinzel}, P., \& {Vial}, J.~C. 1993, \aaps, 99, 513

\bibitem[{Johnston {et~al.}(2019)Johnston, Cargill, Antolin, Hood, De~Moortel,
  \& Bradshaw}]{johnston_effects_2019}
Johnston, C.~D., Cargill, P.~J., Antolin, P., {et~al.} 2019, Astronomy and
  Astrophysics, 625, A149

\bibitem[{Karpen {et~al.}(2006)Karpen, Antiochos, \&
  Klimchuk}]{karpen_origin_2006}
Karpen, J.~T., Antiochos, S.~K., \& Klimchuk, J.~A. 2006, The Astrophysical
  Journal, 637, 531

\bibitem[{{Kawaguchi}(1970)}]{Kawaguchi_1970PASJ...22..405K}
{Kawaguchi}, I. 1970, \pasj, 22, 405

\bibitem[{Klimchuk {et~al.}(2010)Klimchuk, Karpen, \&
  Antiochos}]{klimchuk_can_2010}
Klimchuk, J.~A., Karpen, J.~T., \& Antiochos, S.~K. 2010, \apj, 714, 1239

\bibitem[{Klimchuk \& Luna(2019)}]{klimchuk_role_2019}
Klimchuk, J.~A. \& Luna, M. 2019, The Astrophysical Journal, 884, 68

\bibitem[{{Kohutova} \& {Verwichte}(2016)}]{Kohutova_2016ApJ...827...39K}
{Kohutova}, P. \& {Verwichte}, E. 2016, \apj, 827, 39

\bibitem[{Kohutova {et~al.}(2019)Kohutova, Verwichte, \&
  Froment}]{kohutova_formation_2019}
Kohutova, P., Verwichte, E., \& Froment, C. 2019, Astronomy \& Astrophysics,
  630, A123

\bibitem[{Kuin \& Martens(1982)}]{kuin_thermal_1982}
Kuin, N. P.~M. \& Martens, P. C.~H. 1982, Astronomy and Astrophysics, vol. 108,
  no. 2, Apr. 1982, p. L1-L4. Research supported by the Nederlandse Organisatie
  voor Zuiver-Wetenschappelijk Onderzoek.

\bibitem[{Lemen {et~al.}(2012)Lemen, Title, Akin, Boerner, Chou, Drake, Duncan,
  Edwards, Friedlaender, Heyman, Hurlburt, Katz, Kushner, Levay, Lindgren,
  Mathur, McFeaters, Mitchell, Rehse, Schrijver, Springer, Stern, Tarbell,
  Wuelser, Wolfson, Yanari, Bookbinder, Cheimets, Caldwell, Deluca, Gates,
  Golub, Park, Podgorski, Bush, Scherrer, Gummin, Smith, Auker, Jerram, Pool,
  Soufli, Windt, Beardsley, Clapp, Lang, \& Waltham}]{lemen_atmospheric_2012}
Lemen, J.~R., Title, A.~M., Akin, D.~J., {et~al.} 2012, \solphys, 275, 17

\bibitem[{{Leroy}(1972)}]{Leroy_1972SoPh...25..413L}
{Leroy}, J. 1972, \solphys, 25, 413

\bibitem[{Li {et~al.}(2018)Li, Zhang, Peter, Chitta, Su, Xia, Song, \&
  Hou}]{li_coronal_2018}
Li, L., Zhang, J., Peter, H., {et~al.} 2018, The Astrophysical Journal, 864, L4

\bibitem[{Lionello {et~al.}(2016)Lionello, Alexander, Winebarger, Linker, \&
  Miki{\'c}}]{lionello_can_2016}
Lionello, R., Alexander, C.~E., Winebarger, A.~R., Linker, J.~A., \& Miki{\'c},
  Z. 2016, The Astrophysical Journal, 818, 129

\bibitem[{Lionello {et~al.}(2013)Lionello, Winebarger, Mok, Linker, \&
  Miki{\'c}}]{lionello_thermal_2013}
Lionello, R., Winebarger, A.~R., Mok, Y., Linker, J.~A., \& Miki{\'c}, Z. 2013,
  \apj, 773, 134

\bibitem[{{Liu} {et~al.}(2016){Liu}, {Antolin}, \&
  {Sun}}]{Liu_2016SPD....47.0402L}
{Liu}, W., {Antolin}, P., \& {Sun}, X. 2016, in AAS/Solar Physics Division
  Meeting, Vol.~47, AAS/Solar Physics Division Abstracts \#47, 4.02

\bibitem[{Liu {et~al.}(2012)Liu, Berger, \& Low}]{liu_first_2012}
Liu, W., Berger, T.~E., \& Low, B.~C. 2012, The Astrophysical Journal, 745

\bibitem[{{L{\"o}fdahl}(2002)}]{lofdahl_momfbd_2002SPIE.4792..146L}
{L{\"o}fdahl}, M.~G. 2002, in Presented at the Society of Photo-Optical
  Instrumentation Engineers (SPIE) Conference, Vol. 4792, Society of
  Photo-Optical Instrumentation Engineers (SPIE) Conference Series, ed.
  {P.~J.~Bones, M.~A.~Fiddy, \& R.~P.~Millane}, 146--155

\bibitem[{{Luna} {et~al.}(2012){Luna}, {Karpen}, \&
  {DeVore}}]{Luna_etal_2012ApJ...746...30L}
{Luna}, M., {Karpen}, J.~T., \& {DeVore}, C.~R. 2012, \apj, 746, 30

\bibitem[{Löfdahl {et~al.}(2018)Löfdahl, Hillberg, de~la Cruz~Rodriguez,
  Vissers, Scharmer, Hagfors~Haugan, \& Fredvik}]{lofdahl_data-processing_2018}
Löfdahl, M.~G., Hillberg, T., de~la Cruz~Rodriguez, J., {et~al.} 2018, ArXiv
  e-prints, arXiv:1804.03030

\bibitem[{Mason {et~al.}(2019)Mason, Antiochos, \&
  Viall}]{mason_observations_2019}
Mason, E.~I., Antiochos, S.~K., \& Viall, N.~M. 2019, The Astrophysical
  Journal, 874, L33

\bibitem[{Miki{\'c} {et~al.}(2013)Miki{\'c}, Lionello, Mok, Linker, \&
  Winebarger}]{mikic_importance_2013}
Miki{\'c}, Z., Lionello, R., Mok, Y., Linker, J.~A., \& Winebarger, A.~R. 2013,
  \apj, 773, 94

\bibitem[{Mok {et~al.}(2016)Mok, Miki{\'c}, Lionello, Downs, \&
  Linker}]{mok_three-dimensional_2016}
Mok, Y., Miki{\'c}, Z., Lionello, R., Downs, C., \& Linker, J.~A. 2016, The
  Astrophysical Journal, 817, 15

\bibitem[{Mok {et~al.}(2005)Mok, Miki{\'c}, Lionello, \&
  Linker}]{mok_calculating_2005}
Mok, Y., Miki{\'c}, Z., Lionello, R., \& Linker, J.~A. 2005, The Astrophysical
  Journal, 621, 1098

\bibitem[{{Moschou} {et~al.}(2015){Moschou}, {Keppens}, {Xia}, \&
  {Fang}}]{Moschou_2015AdSpR..56.2738M}
{Moschou}, S.~P., {Keppens}, R., {Xia}, C., \& {Fang}, X. 2015, Advances in
  Space Research, 56, 2738

\bibitem[{{M{\"u}ller} {et~al.}(2003){M{\"u}ller}, {Hansteen}, \&
  {Peter}}]{Muller_2003AA...411..605M}
{M{\"u}ller}, D.~A.~N., {Hansteen}, V.~H., \& {Peter}, H. 2003, \aap, 411, 605

\bibitem[{{M{\"u}ller} {et~al.}(2004){M{\"u}ller}, {Peter}, \&
  {Hansteen}}]{Muller_2004AA...424..289M}
{M{\"u}ller}, D.~A.~N., {Peter}, H., \& {Hansteen}, V.~H. 2004, \aap, 424, 289

\bibitem[{{Neckel}(1999)}]{Neckel1999SoPh..184..421N}
{Neckel}, H. 1999, \solphys, 184, 421

\bibitem[{{Parker}(1953)}]{Parker_1953ApJ...117..431P}
{Parker}, E.~N. 1953, \apj, 117, 431

\bibitem[{Pesnell {et~al.}(2012)Pesnell, Thompson, \&
  Chamberlin}]{pesnell_solar_2012}
Pesnell, W.~D., Thompson, B.~J., \& Chamberlin, P.~C. 2012,
  {\textbackslash}solphys, 275, 3

\bibitem[{{Reep} {et~al.}(2018){Reep}, {Antolin}, \&
  {Bradshaw}}]{reep_electron_2018}
{Reep}, J.~W., {Antolin}, P., \& {Bradshaw}, S.~J. 2018, in AGU Fall Meeting
  Abstracts, Vol. 2018, SH14A--08

\bibitem[{{Schad}(2017)}]{Schad_2017SoPh..292..132S}
{Schad}, T. 2017, \solphys, 292, 132

\bibitem[{Scharmer {et~al.}(2003)Scharmer, Bjelksjo, Korhonen, Lindberg, \&
  Petterson}]{scharmer_1-meter_2003}
Scharmer, G.~B., Bjelksjo, K., Korhonen, T.~K., Lindberg, B., \& Petterson, B.
  2003, Innovative Telescopes and Instrumentation for Solar Astrophysics, 4853,
  341

\bibitem[{{Scharmer} {et~al.}(2003){Scharmer}, {Dettori}, {Lofdahl}, \&
  {Shand}}]{2003SPIE.4853..370S}
{Scharmer}, G.~B., {Dettori}, P.~M., {Lofdahl}, M.~G., \& {Shand}, M. 2003, in
  Society of Photo-Optical Instrumentation Engineers (SPIE) Conference Series,
  Vol. 4853, Society of Photo-Optical Instrumentation Engineers (SPIE)
  Conference Series, ed. S.~L. {Keil} \& S.~V. {Avakyan}, 370--380

\bibitem[{Scharmer {et~al.}(2008)Scharmer, Narayan, Hillberg, de~la
  Cruz~Rodriguez, Löfdahl, Kiselman, Sütterlin, van Noort, \&
  Lagg}]{scharmer_crisp_2008}
Scharmer, G.~B., Narayan, G., Hillberg, T., {et~al.} 2008, The Astrophysical
  Journal Letters, 689, L69

\bibitem[{Schrijver(2001)}]{schrijver_catastrophic_2001}
Schrijver, C.~J. 2001, Solar Physics, 198, 325

\bibitem[{{Scullion} {et~al.}(2016){Scullion}, {Rouppe van der Voort},
  {Antolin}, {Wedemeyer}, {Vissers}, {Kontar}, \&
  {Gallagher}}]{Scullion_2016ApJ...833..184S}
{Scullion}, E., {Rouppe van der Voort}, L., {Antolin}, P., {et~al.} 2016, \apj,
  833, 184

\bibitem[{{Scullion} {et~al.}(2014){Scullion}, {Rouppe van der Voort},
  {Wedemeyer}, \& {Antolin}}]{Scullion_2014ApJ...797...36S}
{Scullion}, E., {Rouppe van der Voort}, L., {Wedemeyer}, S., \& {Antolin}, P.
  2014, \apj, 797, 36

\bibitem[{{Serio} {et~al.}(1981){Serio}, {Peres}, {Vaiana}, {Golub}, \&
  {Rosner}}]{Serio_1981ApJ...243..288S}
{Serio}, S., {Peres}, G., {Vaiana}, G.~S., {Golub}, L., \& {Rosner}, R. 1981,
  \apj, 243, 288

\bibitem[{Testa {et~al.}(2012)Testa, Drake, \& Landi}]{testa_testing_2012}
Testa, P., Drake, J.~J., \& Landi, E. 2012, The Astrophysical Journal, 745, 111

\bibitem[{Ugarte-Urra {et~al.}(2009)Ugarte-Urra, Warren, \&
  Brooks}]{ugarte-urra_active_2009}
Ugarte-Urra, I., Warren, H.~P., \& Brooks, D.~H. 2009, The Astrophysical
  Journal, 695, 642

\bibitem[{Ugarte-Urra {et~al.}(2006)Ugarte-Urra, Winebarger, \&
  Warren}]{ugarte-urra_investigation_2006}
Ugarte-Urra, I., Winebarger, A.~R., \& Warren, H.~P. 2006, The Astrophysical
  Journal, 643, 1245

\bibitem[{van Noort {et~al.}(2005)van Noort, Rouppe van~der Voort, \&
  Löfdahl}]{van_noort_solar_2005}
van Noort, M., Rouppe van~der Voort, L., \& Löfdahl, M.~G. 2005, Solar
  Physics, 228, 191

\bibitem[{Viall \& Klimchuk(2011)}]{viall_patterns_2011}
Viall, N.~M. \& Klimchuk, J.~A. 2011, The Astrophysical Journal, 738, 24

\bibitem[{Viall \& Klimchuk(2012)}]{viall_evidence_2012}
Viall, N.~M. \& Klimchuk, J.~A. 2012, \apj, 753, 35

\bibitem[{Viall \& Klimchuk(2013)}]{viall_modeling_2013}
Viall, N.~M. \& Klimchuk, J.~A. 2013, The Astrophysical Journal, 771, 115

\bibitem[{Viall \& Klimchuk(2015)}]{viall_transition_2015}
Viall, N.~M. \& Klimchuk, J.~A. 2015, The Astrophysical Journal, 799, 58

\bibitem[{Viall \& Klimchuk(2017)}]{viall_survey_2017}
Viall, N.~M. \& Klimchuk, J.~A. 2017, The Astrophysical Journal, 842, 108

\bibitem[{Warren {et~al.}(2002)Warren, Winebarger, \&
  Hamilton}]{warren_hydrodynamic_2002}
Warren, H.~P., Winebarger, A.~R., \& Hamilton, P.~S. 2002, The Astrophysical
  Journal Letters, 579, L41

\bibitem[{{Winebarger} {et~al.}(2018){Winebarger}, {Lionello}, {Downs},
  {Miki{\'c}}, \& {Linker}}]{Winebarger_2018ApJ...865..111W}
{Winebarger}, A.~R., {Lionello}, R., {Downs}, C., {Miki{\'c}}, Z., \& {Linker},
  J. 2018, \apj, 865, 111

\bibitem[{{Winebarger} {et~al.}(2016){Winebarger}, {Lionello}, {Downs},
  {Miki{\'c}}, {Linker}, \& {Mok}}]{Winebarger_2016ApJ...831..172W}
{Winebarger}, A.~R., {Lionello}, R., {Downs}, C., {et~al.} 2016, \apj, 831, 172

\bibitem[{Winebarger {et~al.}(2014)Winebarger, Lionello, Mok, Linker, \&
  Miki{\'c}}]{winebarger_verification_2014}
Winebarger, A.~R., Lionello, R., Mok, Y., Linker, J.~A., \& Miki{\'c}, Z. 2014,
  The Astrophysical Journal, 795, 138

\bibitem[{Xia {et~al.}(2011)Xia, Chen, Keppens, \& van
  Marle}]{xia_formation_2011}
Xia, C., Chen, P.~F., Keppens, R., \& van Marle, A.~J. 2011, The Astrophysical
  Journal, 737, 27

\bibitem[{{Xia} \& {Keppens}(2016)}]{Xia_2016ApJ...823...22X}
{Xia}, C. \& {Keppens}, R. 2016, \apj, 823, 22

\bibitem[{Xia {et~al.}(2014)Xia, Keppens, Antolin, \&
  Porth}]{xia_simulating_2014}
Xia, C., Keppens, R., Antolin, P., \& Porth, O. 2014, The Astrophysical Journal
  Letters, 792, L38

\bibitem[{{Xia} {et~al.}(2017){Xia}, {Keppens}, \&
  {Fang}}]{Xia_2017AA...603A..42X}
{Xia}, C., {Keppens}, R., \& {Fang}, X. 2017, \aap, 603, A42

\end{thebibliography}
\end{document}